# In search of constitutive conditions in isotropic hyperelasticity: polyconvexity versus true-stress-true-strain monotonicity

Maximilian P. Wollner [a,*], Gerhard A. Holzapfel [a,b], Patrizio Neff [c]

[a] Institute of Biomechanics, Graz University of Technology, Stremayrgasse 16/2, 8010, Graz, Austria
[b] Department of Structural Engineering, Norwegian University of Science and Technology, Høgskoleringen 6, 7491, Trondheim, Norway
[c] Chair of Nonlinear Analysis and Modelling, University of Duisburg-Essen, Thea-Leymann-Straße 9, 45127, Essen, Germany

ABSTRACT

The polyconvexity of a strain-energy function is nowadays increasingly presented as the ultimate material stability condition for an idealized elastic response. While the mathematical merits of polyconvexity are clearly understood, its mechanical consequences have received less attention. In this contribution we contrast polyconvexity with the recently rediscovered true-stress-true-strain monotonicity (TSTS-M$^{++}$) condition. By way of explicit examples, we show that neither condition by itself is strong enough to guarantee physically reasonable behavior for ideal isotropic elasticity. In particular, polyconvexity does not imply a monotone trajectory of the Cauchy stress in unconstrained uniaxial extension which TSTS-M$^{++}$ ensures. On the other hand, TSTS-M$^{++}$ does not impose a monotone Cauchy shear stress response in simple shear which is enforced by Legendre-Hadamard ellipticity and in turn polyconvexity. Both scenarios are proven through the construction of appropriate strain-energy functions. Consequently, a combination of polyconvexity, ensuring Legendre-Hadamard ellipticity, and TSTS-M$^{++}$ seems to be a viable solution to Truesdell's Hauptproblem. However, so far no isotropic strain-energy function has been identified that satisfies both constraints globally at the same time. Although we are unable to deliver a valid solution here, we provide several results that could prove helpful in the construction of such an exceptional strain-energy function.

*In memory of* Miroslav Šilhavý (1949–2025)

## 1. Introduction

In the theory of hyperelasticity, the stress response can be derived from a strain-energy (density) function $W$ per reference volume. The search for appropriate constitutive constraints on $W$ has been dubbed the 'Hauptproblem' of finite elasticity by (Truesdell, 1956). Since the second law of thermodynamics has no bearing on the form of the strain-energy function, one needs to find other requirements to ensure physically reasonable material behavior, the notion of which is somewhat arbitrary. In this work, we will focus solely on isotropic solids and an idealized elastic response. By this we do not necessarily mean to describe any real physical material which may show fatigue, failure, softening, local buckling, plasticity, etc. These departures from a perceived ideal elastic response should be incorporated into the modeling framework by additional mechanisms, e.g., a softening law. However, before any such extension, one must agree on this initial elastic idealization. Over the years, several restrictions have been developed in this direction both on the grounds of stability and in an *ad-hoc* manner which we will review in the following.

* Corresponding author.
*E-mail addresses:* wollner@tugraz.at (M.P. Wollner), holzapfel@tugraz.at (G.A. Holzapfel), patrizio.neff@uni-due.de (P. Neff).

One approach is to constrain the material response to disturbances from some stable state of deformation. One such statement is given by quasiconvexity which requires that a spatially homogeneous, hyperelastic body, defined in its reference configuration over $\Omega$ and constrained at the boundary $\partial\Omega$, attains its minimal strain-energy for a homogeneous deformation, i.e.,

$$\int_\Omega W(\mathbf{F}+\boldsymbol{\nabla}\boldsymbol{\vartheta})\,\mathrm{d}V \geq \int_\Omega W(\mathbf{F})\,\mathrm{d}V = \mathrm{vol}(\Omega)\,W(\mathbf{F}), \tag{1.1}$$

cf. (Morrey, 1952) and (Šilhavý, 1997, Eq. (17.1.3)). Here, $\mathbf{F}\in\mathrm{GL}^+(3)$ is a constant deformation gradient, while $\boldsymbol{\nabla}\boldsymbol{\vartheta}$ is the displacement gradient of some disturbance which vanishes on the boundary of $\Omega$, i.e., $\boldsymbol{\vartheta}(\boldsymbol{X})=\boldsymbol{0}\,\forall\boldsymbol{X}\in\partial\Omega$.[1] The condition is also intimately linked to existence proofs in non-linear elasticity, cf. (Ball, 1976). Given the integral nature of quasiconvexity, the condition is difficult to prescribe *a priori*. Therefore, one frequently resorts to the stricter requirement of polyconvexity which ensures quasiconvexity and is considerably easier to handle. To this end, one introduces some convex function $\mathcal{P}(\mathbf{F},\mathbf{G},\delta)$ and sets $W(\mathbf{F})=\mathcal{P}(\mathbf{F},\mathrm{Cof}\,\mathbf{F},\det\mathbf{F})$, such that

$$W(\overline{\mathbf{F}}) \geq W(\mathbf{F}) + \Big\langle \frac{\partial\mathcal{P}}{\partial\mathbf{F}}\Big|_{\mathbf{F}}, \overline{\mathbf{F}}-\mathbf{F}\Big\rangle + \Big\langle \frac{\partial\mathcal{P}}{\partial\mathbf{G}}\Big|_{\mathrm{Cof}\,\mathbf{F}}, \mathrm{Cof}\,\overline{\mathbf{F}}-\mathrm{Cof}\,\mathbf{F}\Big\rangle + \frac{\partial\mathcal{P}}{\partial\delta}\Big|_{\det\mathbf{F}}(\det\overline{\mathbf{F}}-\det\mathbf{F}) \quad \forall\,\mathbf{F},\overline{\mathbf{F}}\in\mathrm{GL}^+(3). \tag{1.2}$$

By taking $\overline{\mathbf{F}}=\mathbf{F}+\boldsymbol{\nabla}\boldsymbol{\vartheta}$ and integrating over $\Omega$, quasiconvexity follows directly, cf. (Krawietz, 1986, Eqs. (12.91)–(12.96)). Notably, polyconvexity itself does not have a direct physical or mechanical interpretation beyond its implication of quasiconvexity. It is interesting though that the proof for polyconvexity uses the fact that the volumetric averages of line, area, and volume elements remain unaffected by the superposed fluctuation $\boldsymbol{\vartheta}$. At any rate, polyconvexity can always be treated as a mathematical convenience.

Another constitutive constraint – implied by quasiconvexity and in turn by polyconvexity – is rank-one convexity, cf. (Šilhavý, 1997, Sect. 17.3). Here,

$$W(\mathbf{F}+t\,\boldsymbol{a}\otimes\boldsymbol{b}) \leq t\,W(\mathbf{F}+\boldsymbol{a}\otimes\boldsymbol{b}) + (1-t)W(\mathbf{F}) \quad \forall t\in[0,1] \quad \forall\,\mathbf{F},\mathbf{F}+\boldsymbol{a}\otimes\boldsymbol{b}\in\mathrm{GL}^+(3) \quad \forall\boldsymbol{a},\boldsymbol{b}\in\mathbb{R}^3 \tag{1.3}$$

or, given sufficient differentiability, the Legendre-Hadamard condition

$$\langle \mathrm{D}^2_{\mathbf{F}}W(\mathbf{F}).(\boldsymbol{a}\otimes\boldsymbol{b}), \boldsymbol{a}\otimes\boldsymbol{b}\rangle \geq 0. \tag{1.4}$$

Physically, the rank-one convexity ensures infinitesimal stability against interior perturbations and, in its strict form, real wave speeds in incremental elastic deformations, cf. (Truesdell and Noll, 1965, Sects. 68 bis. & 71). Notably, (Bertram et al., 2007) showed that a strain-energy function leading to a physically linear constitutive relation in some generalized Seth-Hill strain measure cannot be rank-one convex. The generalization of rank-one convexity to convexity directly in $\mathbf{F}$ is incompatible with physical requirements such as $\lim_{\det\mathbf{F}\to0^+}W(\mathbf{F})=\infty$, cf. (Ciarlet, 1988, Sect. 4.8), or the non-uniqueness of solutions, cf. [Chap. 10](Bigoni, 2012).

Necessary and sufficient conditions for rank-one convexity in three dimensions in terms of principal stretches are given by (Aubert, 1988, Theo. 4.2). Sufficient conditions for polyconvexity have been found by (Ball, 1976, Theo. 5.2) and (Rosakis, 1997, Theo. 3.1), while (Mielke, 2005, Theo. 2.2) also provides necessary ones, albeit in a form difficult to apply. A transition from principal stretches to signed singular values considerably simplifies the representation of these necessary and sufficient conditions, cf. (Wiedemann and Peter, 2025, Theo. 2.1).

Another class of constitutive constraints is related to the monotonicity between different stress and strain measures. The inherent ambiguity of defining a strain tensor at finite deformations has lead to a host of possibilities, cf. (Truesdell and Toupin, 1960, Sect. 33) and (Liu et al., 2026, Sect. 2.1). While there is no correct choice in a strict sense, not all strain measures exhibit the same mathematical properties. For example, the Eulerian strain measure $\mathbf{V}-\mathbb{1}$, where $\mathbf{V}$ denotes the left stretch tensor, diverges for large volumetric expansion, but converges to $-\mathbb{1}$ for large volumetric contraction. In contrast, other functional choices such as the Eulerian logarithmic strain $\log\mathbf{V}$ uphold coercivity.[2] While the parametrization of a strain-energy function in terms of a specific strain measure is largely irrelevant for the purpose of deriving a stress response, it does matter for the definition of monotonicity constraints in three-dimensions. Take for example the two monotonicity statements between either $\mathbf{V}-\mathbb{1}$ or $\log\mathbf{V}$ and the Cauchy (true) stress $\boldsymbol{\sigma}$ such that

$$\langle\overline{\boldsymbol{\sigma}}-\boldsymbol{\sigma}, \overline{\mathbf{V}}-\mathbf{V}\rangle > 0 \quad \forall\,\mathbf{V},\overline{\mathbf{V}}\in\mathrm{Sym}^{++}(3),\ \mathbf{V}\neq\overline{\mathbf{V}}, \tag{1.5}$$

$$\langle\overline{\boldsymbol{\sigma}}-\boldsymbol{\sigma}, \log\overline{\mathbf{V}}-\log\mathbf{V}\rangle > 0 \quad \forall\,\mathbf{V},\overline{\mathbf{V}}\in\mathrm{Sym}^{++}(3),\ \mathbf{V}\neq\overline{\mathbf{V}}. \tag{1.6}$$

The two barred and non-barred quantities each constitute a stress-strain pair. It is not immediately obvious if one implies the other or whether these are independent requirements. Indeed, the two candidates coincide in the one-dimensional case, but not generally in the multiaxial case.

Since the specific choice of a pair of stress and strain measures is not necessarily mandated by some deeper underlying concept, these types of inequalities are taken *a priori*, cf., (Krawietz, 1975), (Šilhavý, 1997, Sect. 18.6), and (Ghiba et al., 2025). Due to the ensuing range of possibilities, one can come up with a whole hierarchy of constraints, cf. (Truesdell and Noll, 1965, Sects. 51–53) for a summary prior to 1965 and (Neff et al., 2015a, Sect. 2) and (Mihai and Goriely, 2017) for more recent reviews. Particularly noteworthy here is a family of constraints proposed by (Hill, 1968, 1970) which reads

$$\left\langle \frac{\mathrm{D}^{\mathrm{ZJ}}\boldsymbol{\tau}}{\mathrm{D}t} - m\,\boldsymbol{\tau}\,\mathbf{D} - m\,\mathbf{D}\,\boldsymbol{\tau}, \mathbf{D} \right\rangle > 0 \quad \forall \mathbf{F} \in \mathrm{GL}^+(3) \quad \forall \dot{\mathbf{F}} \in \mathbb{R}^{3\times 3}, \tag{1.7}$$

where $\frac{\mathrm{D}^{\mathrm{ZJ}}\boldsymbol{\tau}}{\mathrm{D}t} = \dot{\boldsymbol{\tau}} + \boldsymbol{\tau}\,\mathbf{W} - \mathbf{W}\,\boldsymbol{\tau}$ is the (corotational) Zaremba-Jaumann rate of the Kirchhoff stress $\boldsymbol{\tau}$. The tensors $\mathbf{D}$ and $\mathbf{W}$ denote the symmetric and skew-symmetric parts of the rate of deformation tensor $\mathbf{L} = \dot{\mathbf{F}}\,\mathbf{F}^{-1}$, respectively. The real scalar $m$ is related to the family of generalized Seth-Hill strain tensors, cf. (Seth, 1962, Sect. 2) and (Hill, 1968, Eq. (3)). Interestingly, for the choice $m = \frac{1}{2}$, one recovers a stricter version of the Coleman-Noll condition, cf. (Coleman and Noll, 1959, Eq. (8.8)), which was considered one possible solution to the 'Hauptproblem' at the time. This can be seen by (Coleman and Noll, 1964, Theo. 2), while remembering that $\boldsymbol{\tau} = J\,\boldsymbol{\sigma}$ with $J = \det \mathbf{F}$. Indeed, (Hill, 1968, Eq. (30)) rejects any $m \neq 0$ and therefore the Coleman-Noll condition due to physical inconsistencies arising by incorporating incompressibility. We refer to the particular choice $m = 0$ as Hill's inequality. A preference for $m = 0$ is also apparent in the work on compressible elastic solids by (Ogden, 1970, Sect. 4). For this choice, the inequality (1.7) implies a monotonicity between the Hencky strain measure $\log \mathbf{V}$ and the Kirchhoff stress $\boldsymbol{\tau}$ with

$$\langle \overline{\boldsymbol{\tau}} - \boldsymbol{\tau}, \log \overline{\mathbf{V}} - \log \mathbf{V} \rangle > 0 \quad \forall \mathbf{V}, \overline{\mathbf{V}} \in \mathrm{Sym}^{++}(3),\ \mathbf{V} \neq \overline{\mathbf{V}}, \tag{1.8}$$

cf. (Hill, 1968, Sect. 4). Since[3]

$$\boldsymbol{\tau} = \mathrm{D}_{\log \mathbf{V}} \widehat{W}(\log \mathbf{V}) \qquad \text{with} \qquad W(\mathbf{F}) = \widehat{W}(\log \mathbf{V}), \tag{1.9}$$

it follows that Hill's inequality is satisfied if and only if $\widehat{W}$ is convex in the Hencky strain $\log \mathbf{V}$, cf. (Hill, 1970, Sect. 3).

[1] A more detailed explanation of the notation and the basic quantities is given in Appendix A and Sect. 2, respectively.

[2] Early usage of a logarithmic measure can be found in (Imbert, 1880, p. 53), (Becker, 1893, Eq. (5)), and (Ludwik, 1909, p. 17). Other common names for the logarithmic strain include natural strain, true strain, and Hencky strain, cf. (Freed, 1997). For the majority of this work, we will stick with the latter.

It should be noted that the argument by (Hill, 1968) based on the incompressiblity constraint has been rejected by (Wang and Truesdell, 1973, p. 235–238), which in turn has been heavily criticized by (Rivlin, 1973, Sect. 12.6) and again in (Rivlin, 2004). There is however another objection to (1.8) as it entails – for a perfect fluid with mass density per current volume $\rho$ and the constitutive relation $\boldsymbol{\sigma} = -p(\rho)\mathbb{1}$ – the constraint

$$\frac{\mathrm{d}p}{\mathrm{d}\rho} > \frac{p}{\rho}, \tag{1.10}$$

which is overly restrictive for 'a fluid capable of change of phase', cf. (Wang and Truesdell, 1973, p. 258); see also (Šilhavý, 1997, Sect. 19). There is also another illustrative representation of inequality (1.10). In case the pressure of the perfect fluid can be derived from a strain-energy function, we have $W(\mathbf{F}) = h(J)$ and $p = -\frac{\mathrm{d}h}{\mathrm{d}J}$ such that

$$\frac{\mathrm{d}p}{\mathrm{d}\rho} > \frac{p}{\rho} \quad \Longleftrightarrow \quad J\,\frac{\mathrm{d}^2 h}{\mathrm{d}J^2} + \frac{\mathrm{d}h}{\mathrm{d}J} > 0 \quad \Longleftrightarrow \quad h \text{ is strictly convex in } \log J. \tag{1.11}$$

Hence, strict convexity of $h$ in $J$ alone is not enough to ensure Hill's inequality.

Since (1.7) with $m = 0$ performs well for incompressible materials, it is a natural next step to analyze the constitutive inequality

$$\left\langle \frac{\mathrm{D}^{\mathrm{ZJ}}\boldsymbol{\sigma}}{\mathrm{D}t}, \mathbf{D} \right\rangle > 0 \quad \forall \mathbf{F} \in \mathrm{GL}^+(3) \quad \forall \dot{\mathbf{F}} \in \mathbb{R}^{3\times 3}, \tag{1.12}$$

i.e., replacing the Kirchhoff stress $\boldsymbol{\tau}$ with the Cauchy stress $\boldsymbol{\sigma}$. This task was taken up by (Leblond, 1992) for hyperelastic materials. After several explicit examples, Leblond comes to the conclusion that the use of the Zaremba-Jaumann rate of the Cauchy stress is superior to the Kirchhoff stress. Here, we retrieve the classic constraint $\frac{\mathrm{d}p}{\mathrm{d}\rho} > 0$ for a perfect fluid, cf. (Truesdell, 1980, Eq. (2A.6)). In contrast to (1.11), the inequality (1.12) then corresponds to $h$ being necessarily strictly convex in $J$ making it virtually identical to polyconvexity in case of a perfect fluid, cf. (Leblond, 1992, Eq. (9)). For incompressible solids, the inequality (1.12) reduces to Hill's inequality. In case of hyperelasticity, necessary and sufficient conditions for (1.12) in terms of principal stretches are already provided in the original paper, cf. (Leblond, 1992, Eq. (23)). The more general follow-up work by (d'Agostino et al., 2025, Rem. A.8) for Cauchy elasticity establishes that

$$\left\langle \frac{\mathrm{D}^{\mathrm{ZJ}}\boldsymbol{\sigma}}{\mathrm{D}t}, \mathbf{D} \right\rangle > 0 \quad \Longleftrightarrow \quad \left\langle \mathrm{D}_{\log \mathbf{V}} \widehat{\boldsymbol{\sigma}}(\log \mathbf{V}).\mathbf{H}, \mathbf{H} \right\rangle > 0 \quad \forall \mathbf{V} \in \mathrm{Sym}^{++}(3) \quad \forall \mathbf{H} \in \mathrm{Sym}(3) \setminus \{\mathbf{0}\} \quad (\text{TSTS-M}^{++}) \tag{1.13}$$

$$\Longrightarrow \quad \langle \overline{\boldsymbol{\sigma}} - \boldsymbol{\sigma}, \log \overline{\mathbf{V}} - \log \mathbf{V} \rangle > 0 \quad \forall \mathbf{V}, \overline{\mathbf{V}} \in \mathrm{Sym}^{++}(3),\ \mathbf{V} \neq \overline{\mathbf{V}}, \qquad (\text{TSTS-M}^{+}) \tag{1.14}$$

i.e., a hierarchy of constraints related to **t**rue-**s**tress-**t**rue-**s**train **m**onotonicity; here, $\boldsymbol{\sigma} = \widehat{\boldsymbol{\sigma}}(\log \mathbf{V})$. It has been shown in (Neff et al., 2025d) that TSTS-M$^{++}$ implies positive incremental Cauchy stress moduli for spatially homogeneous, diagonal deformations. TSTS-M$^{++}$ might also provide a pathway to proving the local existence of solutions in finite nonlinear isotropic elasticity, cf. (Neff et al., 2026). Interestingly, TSTS-M$^{++}$ has also been used independently by (Jog and Patil, 2013) to identify material instabilities.

[3] Although we can already see glimpses of the fact that the Kirchhoff stress $\boldsymbol{\tau}$ and the Hencky strain $\log \mathbf{V}$ constitute a conjugate pair in isotropic hyperelasticity in (Hencky, 1929), the relation is – to the knowledge of the authors – first made explicit by (Murnaghan, 1941, p. 127). Here, we find in the original nomenclature $\mathbf{N} = \exp(-2\mathbf{R})$ and $\mathbf{T} = \varrho \frac{\partial \varphi}{\partial \mathbf{R}}$, where $\mathbf{T}$ denotes the Cauchy stress, $\varrho$ the current mass density, and $\varphi$ an elastic energy per unit mass. From (Murnaghan, 1941, p. 122), we can see that $\mathbf{N} = \mathbf{Q}^{\mathrm{T}}\mathbf{Q}$, where $\mathbf{Q}$ denotes the inverse deformation gradient and with (Murnaghan, 1941, p. 129) we have $\varrho = \varrho_0 \det \mathbf{Q}$, where $\varrho_0$ denotes the mass density with respect to the reference volume. Converting all this into our notation, we have $\mathbf{Q} = \mathbf{F}^{-1}$, $\mathbf{N} = \mathbf{B}^{-1}$, $\mathbf{R} = \log \mathbf{V}$, $\rho = \frac{\rho_0}{J}$, $\varphi = \frac{\widehat{W}}{\rho_0}$, and $\mathbf{T} = \frac{1}{J}\boldsymbol{\tau}$. Consequently, the relation (1.9) follows. Said expression can also be found later in (Richter, 1948, Eq. (3.8*)), cf. (Graban et al., 2019). Richter was most likely unaware of Murnaghan's work.

A valid criticism of Leblond's inequality and in turn TSTS-M$^{++}$ is the specific choice of the Zaremba-Jaumann rate. It is therefore reasonable to ask whether the equivalence (1.13) also hold for any other objective rate '∘', i.e.,

$$\left\langle \frac{\mathrm{D}^\circ \boldsymbol{\sigma}}{\mathrm{D}t}, \mathbf{D} \right\rangle > 0 \quad \forall\, \mathbf{F} \in \mathrm{GL}^+(3) \quad \forall\, \dot{\mathbf{F}} \in \mathbb{R}^{3\times 3} \quad \overset{(?)}{\iff} \quad \text{TSTS-M}^{++}. \tag{1.15}$$

Indeed, it has been shown in (Neff et al., 2025a, Prop. 4.10) that the statement also holds for the corotional logarithmic rate. In general, an equivalence of such a **c**orotational **s**tability **p**ostulate (CSP) and TSTS-M$^{++}$ would further support to the importance of the latter. We conjecture that the equivalence holds for all 'reasonable' corotational rates, cf. (Neff et al., 2025c), which is to be shown in an upcoming publication.

Although (Leblond, 1992, p. 463) remarks that '*a thorough investigation* [of (1.12)] *would be worthwhile*', comparatively little is still known about its physical consequences. As shown by (Leblond, 1992, Sect. 4b), TSTS-M$^{++}$ does in general not entail polyconvexity and *vice-versa*. In response, (Neff et al., 2024) have recently put forward several challenge questions that try to elucidate the interaction of TSTS-M$^{++}$ and polyconvexity in physically relevant deformation modes such as unconstrained uniaxial extension-compression and simple shear at large strains. Four of these five read as follows:

(i) **Combination of polyconvexity and TSTS-M$^{++}$**: Find a compressible strain-energy function $W$ that is polyconvex (or rank-one convex) and satisfies TSTS-M$^{++}$ globally for all $\mathbf{F} \in \mathrm{GL}^+(3)$. The resulting constitutive relation for the Cauchy stress must be bijective and must linearize to a proper elastic law in the infinitesimal theory.

(ii) **Insufficiency of polyconvexity (compressible):** Find a compressible strain-energy function $W$ that is polyconvex (or rank-one convex), that shows a non-monotonic true-stress response in unconstrained uniaxial extension-compression, and that linearizes to a proper elastic law in the infinitesimal theory.

(iii) **Insufficiency of polyconvexity (incompressible):** Find an incompressible strain-energy function $W$ that is polyconvex (or rank-one convex), that shows a non-monotonic true-stress response in unconstrained uniaxial extension-compression, and that linearizes to a proper elastic law in the infinitesimal theory.

(iv) **Insufficiency of TSTS-M$^{++}$**: Find a compressible strain-energy function $W$ that satisfies TSTS-M$^{++}$, that shows a non-monotonic true-shear-stress response in simple shear, and that linearizes to a proper elastic law in the infinitesimal theory.

Alternatively, show that any such $W$ is impossible.[4]

If an elastic constitutive relation adheres to TSTS-M$^{++}$, then by implication (1.14) the Cauchy stress response in unconstrained uniaxial extension-compression is strictly monotonic. This can straightforwardly be seen with

$$\langle \overline{\boldsymbol{\sigma}} - \boldsymbol{\sigma}, \log \overline{\mathbf{V}} - \log \mathbf{V} \rangle > 0 \quad \Longrightarrow \quad (\overline{\sigma}_{11} - \sigma_{11})(\log \overline{\lambda}_1 - \log \lambda_1) > 0 \quad \Longrightarrow \quad \overline{\sigma}_{11} > \sigma_{11} \quad \text{if} \quad \overline{\lambda}_1 > \lambda_1, \tag{1.16}$$

where $\sigma_{11}$ is the only non-zero component of $\sigma$ by definition of the boundary value problem for an applied stretch $\lambda_1$. It is unclear whether polyconvexity guarantees such behavior, hence Challenge (ii) and (iii). On the other hand, polyconvexity implies rank-one convexity which in turn ensures a monotonic true-shear-stress response in simple shear. A proof can be found following Prop. 5.13 in this contribution. Whether TSTS-M$^{++}$ implies such a constraint is not immediately obvious, hence Challenge (iv). An overview of these implications is given in Fig. 1.

In this work, we will provide full solutions to Challenges (ii) and (iv) by constructing an appropriate family of strain-energy functions. Notably, in (Korobeynikov et al., 2025, Sect 6.2.3) a solution to Challenge (ii) has already been given in unconstrained uniaxial compression. We instead provide a solution in extension. Consequently, polyconvexity alone is not sufficient to guarantee a physically meaningful material response. This might be especially relevant for constitutive neural networks, where polyconvexity is often the sole constitutive constraint considered in this respect, setting aside such obvious requirements as objectivity, cf. (Klein et al., 2022; Linka and Kuhl, 2023; Linden et al., 2023; Geuken et al., 2025).

For the remaining two challenges, we can only provide partial results. For Challenge (i), we construct three families of strain-energy functions that satisfy both polyconvexity and TSTS-M$^{++}$, albeit in a chain-limited setting, i.e., not globally defined as required. In case of Challenge (iii), we show that an incompressible strain-energy function that satisfies the sufficient condition of polyconvexity by (Ball, 1976, Theo. 5.2) automatically leads to a monotonic true-stress response in unconstrained uniaxial extension-compression. This is obviously not enough to show the impossibility of a solution to Challenge (iii), but it seriously reduces the space of candidates. Besides tackling these specific questions, we also provide several general results related to polyconvexity and TSTS-M$^{++}$ which have – to the knowledge of the authors – not yet been discussed in the literature. None of the proofs in this work resort to large-scale computation, except for visualization purposes or to speed up the tedious task of linearization through symbolic differentiation.

Concerning the structure of this work, we briefly introduce all relevant mathematical quantities and relations in Sect. 2. Since a theorem is often only half as interesting as its proof, we provide several results related to sufficient conditions for polyconvexity and TSTS-M$^{++}$ in Sects. 3 and 4, respectively, which are subsequently used to (partially) answer the challenge questions in Sect. 5. We conclude with a short summary and outlook in Sect. 6.

## 2. Isotropic hyperelasticity

Each material point, initially located at $\boldsymbol{X} \in \mathbb{R}^3$, is assigned its current coordinates $\boldsymbol{x} \in \mathbb{R}^3$ through some motion $\boldsymbol{x} = \boldsymbol{\varphi}(\boldsymbol{X}, t)$. The deformation gradient is defined as $\mathbf{F} = \boldsymbol{\nabla}\boldsymbol{\varphi} \in \mathrm{GL}^+(3)$ with positive determinant $J = \det \mathbf{F} > 0$. The left Cauchy-Green tensors and left stretch tensor follow with $\mathbf{B} = \mathbf{F}\mathbf{F}^{\mathrm{T}}$ and $\mathbf{V} = \sqrt{\mathbf{B}}$, respectively, cf. (Holzapfel, 2000, Chap. 2). The two foregoing tensors are all elements of $\mathrm{Sym}^{++}(3)$.

[4] For the solution of Challenge (i), Patrizio Neff is offering a prize money of 500€.

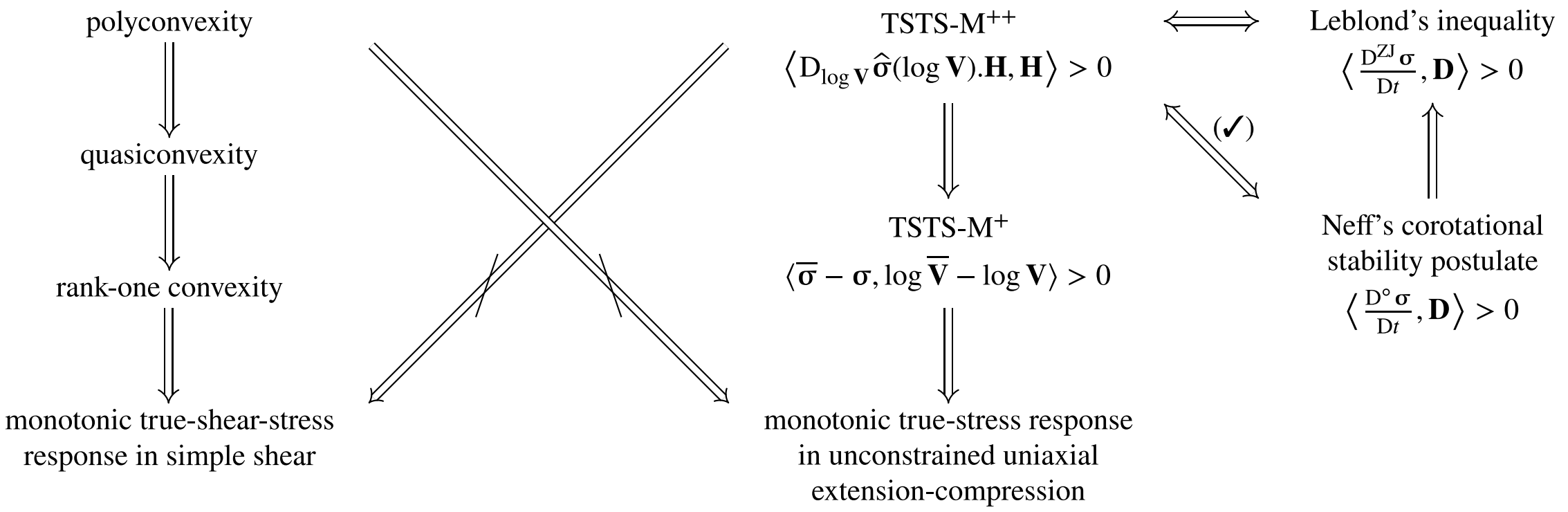

**Fig. 1.** A visual overview of the various constitutive constraints and their implications. As shown by construction in Sects. 5.2 and 5.3, polyconvexity and TSTS-M$^{++}$ do not imply monotonicity in unconstrained uniaxial extension-compression and simple shear, respectively.

The three principal invariants of $\mathbf{B}$ read

$$I_1 = \operatorname{tr} \mathbf{B} = \|\mathbf{F}\|^2, \qquad I_2 = \frac{1}{2}\left((\operatorname{tr} \mathbf{B})^2 - \operatorname{tr} \mathbf{B}^2\right) = \|\operatorname{Cof} \mathbf{F}\|^2, \qquad \text{and} \qquad I_3 = \det \mathbf{B} = (\det \mathbf{F})^2, \tag{2.1}$$

cf. (Ogden, 1997, Sect. 1.3.2). We will however mainly use an alternative set of invariants $K_i$ defined as the square roots of $I_i$ which leads to simpler representation of constitutive inequalities. Hence,

$$K_1 = \sqrt{I_1} = \|\mathbf{F}\|, \qquad K_2 = \sqrt{I_2} = \|\operatorname{Cof} \mathbf{F}\|, \qquad \text{and} \qquad K_3 = \sqrt{I_3} = \det \mathbf{F}. \tag{2.2}$$

Notably $\mathbf{F}$, $\operatorname{Cof} \mathbf{F}$, and $\det \mathbf{F}$ capture information about the deformation of an line, area, and volume element, respectively, cf. (Kearsley, 1989) and (Wollner et al., 2023, Sect. 3).

The left stretch tensor $\mathbf{V}$ allows for a spectral decomposition with

$$\mathbf{V} = \sum_{i=1}^{3} \lambda_i \, \boldsymbol{v}_i \otimes \boldsymbol{v}_i, \tag{2.3}$$

where $\lambda_i$ denote three distinct principal stretches and $\boldsymbol{v}_i$ the associated principal direction, cf. (Šilhavý, 1997, Sect. 1.2.1). For three distinct principal stretches, the right-hand orthonormal system of eigenvectors is unique up to a $\pi$-rotation around any of the base vectors. In the case of repeated eigenvalues, the orthonormal system is no longer uniquely defined reflecting the increased symmetry. We define the Hencky strain measure $\log \mathbf{V} \in \mathrm{Sym}(3)$ with

$$\log \mathbf{V} = \sum_{i=1}^{3} \log(\lambda_i) \, \boldsymbol{v}_i \otimes \boldsymbol{v}_i. \tag{2.4}$$

Throughout this work, we assume the existence of an isotropic, spatially homogeneous, continuous strain-energy (density) function $W$ per unit reference volume. Due to objectivity and material symmetry, the function must be representable through the invariants $I_i$ and in turn $K_i$, i.e., $W(\mathbf{F}) = \Psi(K_i)$, cf. (Truesdell and Noll, 1965, Sect. 85). In case of isotropic hyperelasticity, we can compute the Cauchy (true) stress $\boldsymbol{\sigma}$ from

$$\boldsymbol{\sigma} = \frac{1}{J} \mathrm{D}_{\log \mathbf{V}} \widehat{W}(\log \mathbf{V}) = \frac{1}{K_3} \sum_{i=1}^{3} \frac{\partial \Psi}{\partial K_i} \mathrm{D}_{\log \mathbf{V}} K_i, \tag{2.5}$$

which follows from the conjugate properties of the Kirchhoff stress $\boldsymbol{\tau}$ and the Hencky strain $\log \mathbf{V}$, cf. (Murnaghan, 1941; Richter, 1948; Hill, 1968).

Representing the invariants (2.2) in terms of $\log \mathbf{V}$ reads[5]

$$K_1 = \sqrt{\operatorname{tr} \exp(2 \log \mathbf{V})}, \qquad K_2 = \exp(\operatorname{tr} \log \mathbf{V}) \sqrt{\operatorname{tr} \exp(-2 \log \mathbf{V})}, \qquad \text{and} \qquad K_3 = \exp(\operatorname{tr} \log \mathbf{V}) \tag{2.6}$$

with the tensor derivatives

$$\mathrm{D}_{\log \mathbf{V}} K_1 = \frac{1}{2K_1} \mathrm{D}_{\log \mathbf{V}} \left(\operatorname{tr} \exp(2 \log \mathbf{V})\right) = \frac{\exp(2 \log \mathbf{V})}{K_1} = \frac{\mathbf{B}}{K_1}, \tag{2.7}$$

$$\mathrm{D}_{\log \mathbf{V}} K_2 = \mathrm{D}_{\log \mathbf{V}} \left(\exp(\operatorname{tr} \log \mathbf{V})\right) \sqrt{\operatorname{tr} \exp(-2 \log \mathbf{V})} + \frac{\exp(\operatorname{tr} \log \mathbf{V})}{2\sqrt{\operatorname{tr} \exp(-2 \log \mathbf{V})}} \mathrm{D}_{\log \mathbf{V}} \left( \operatorname{tr} \exp(-2 \log \mathbf{V})\right)$$

[5] The exponential function with a second-order symmetric tensor as an argument is treated analogously to the tensor logarithm in (2.4), cf. (Šilhavý, 1997, Sect. 8.1.5).

$$= \Big(\exp(\operatorname{tr}\log \mathbf{V})\sqrt{\operatorname{tr}\exp(-2\log \mathbf{V})}\Big)\mathbb{1} - \frac{\exp(\operatorname{tr}\log \mathbf{V})\exp(-2\log \mathbf{V})}{\sqrt{\operatorname{tr}\exp(-2\log \mathbf{V})}}$$

$$= K_2\mathbb{1} - \frac{\exp(2\operatorname{tr}\log \mathbf{V})\exp(-2\log \mathbf{V})}{K_2} = \frac{K_2^2\mathbb{1} - \operatorname{Cof}\mathbf{B}}{K_2}, \tag{2.8}$$

$$\mathrm{D}_{\log \mathbf{V}} K_3 = \mathrm{D}_{\log \mathbf{V}}\big(\exp(\operatorname{tr}\log \mathbf{V})\big) = K_3\mathbb{1}. \tag{2.9}$$

In the undeformed configuration $\mathbf{F} = \mathbb{1}$, the stress must vanish which leads to the additional scalar constraint

$$\left(\frac{\partial \Psi}{\partial K_1} + 2\frac{\partial \Psi}{\partial K_2} + \sqrt{3}\frac{\partial \Psi}{\partial K_3}\right)\bigg|_{\mathbf{B}=\mathbb{1}} = 0. \tag{2.10}$$

As an alternative to invariants, we can represent the strain-energy function in terms of the principal stretches, i.e., $W(\mathbf{F}) = \psi(\lambda_1, \lambda_2, \lambda_3)$, where $\psi$ obeys a permutation invariance with respect to its arguments. The Cauchy stress follows with

$$\sigma = \frac{1}{J}\mathrm{D}_{\log \mathbf{V}}\widehat{W}(\log \mathbf{V}) = \frac{1}{\lambda_1\lambda_2\lambda_3}\sum_{i=1}^{3}\lambda_i\frac{\partial \psi}{\partial \lambda_i}\boldsymbol{v}_i \otimes \boldsymbol{v}_i, \tag{2.11}$$

cf. (Ogden, 1997, Sect. 4.3.4).

In the classical infinitesimal theory of isotropic elasticity, which can be seen as a first order approximation of any isotropic elastic law at small strains around a stress-free reference state, the material behavior is fully defined by two Lamé constants $\lambda$ and $\mu$.[6] Consequently, we can derive these two constants by linearization of (2.5), although the expressions can quickly become unwieldy. An efficient approach is presented in (Truesdell and Noll, 1965, Eq. (50.13)) which is readily implemented in a software environment capable of symbolic differentiation, e.g., Mathematica (Wolfram Research, 2023). For our purposes, a proper elastic law in the infinitesimal theory requires that

$$\mu > 0 \quad \text{and} \quad 2\mu + 3\lambda > 0, \tag{2.12}$$

cf. (Truesdell and Noll, 1965, Eq. (51.1)). These conditions are necessary and sufficient for the strict convexity of the strain-energy function in the infinitesimal theory. An elastic response function that satisfies TSTS-M$^{++}$ automatically fulfills the requirement 2.12, which can be easily seen by linearizing (1.13), cf. (Leblond, 1992, p. 450). The condition of polyconvexity in the infinitesimal theory on the other hand does not enforce (2.12), but instead implies only

$$\mu \geq 0 \quad \text{and} \quad 2\mu + \lambda \geq 0, \tag{2.13}$$

cf. (Krawietz, 1986, Sect. 12.5) and (Leblond, 1992, App. B).

While the shear modulus $\mu$ has a straightforward physical interpretation, the first Lamé constant is better understood through its relation to the bulk modulus $\kappa$ and Poisson's ratio $\nu$ defined by

$$\kappa = \frac{2\mu + 3\lambda}{3} \quad \text{and} \quad \nu = \frac{1}{2}\frac{\lambda}{\lambda + \mu}, \tag{2.14}$$

respectively, cf. (Truesdell and Noll, 1965, Sect. 51) and (Ogden, 1997, Sect. 6.1.6).

In case of incompressibility, the strain-energy function $W$ only needs to be defined for isochoric deformations states, i.e., $J = 1$. In the elastic response function, this additional constraint introduces a Lagrange parameters $p$, such that

$$\sigma = -p\mathbb{1} + \mathrm{D}_{\log \mathbf{V}}\widehat{W}(\log \mathbf{V}), \tag{2.15}$$

cf. (Truesdell and Noll, 1965, Sect. 30) and (Ogden, 1997, Sect. 4.3.5). In case of incompressibility and isotropy, the requirement of a stress-free initial configuration (2.10) is trivially fulfilled for an appropriate choice of $p$. In correspondence with the infinitesimal theory, there only remains the shear modulus $\mu$ which can be calculated according to (Truesdell and Noll, 1965, Eq. (50.14)).

**Remark 2.1.** *Here, we want to highlight some potentially lesser known instances for the usage of the Hencky strain in the history of elastic constitutive modeling. Although this particular strain measure has been deemed by some impractical for its algebraic complexity, cf. (Truesdell and Toupin, 1960, Sect 33), we may find usage of* $\log \mathbf{V}$ *as early as (Becker, 1893). In a modern interpretation of Becker's work, we have*

$$\sigma = \frac{1}{J}\big(2\mu\log \mathbf{V} + \lambda\operatorname{tr}(\log \mathbf{V})\mathbb{1}\big)\mathbf{V}, \tag{2.16}$$

*cf. (Neff et al., 2016b, Sect. 1.2). Other early appearances of the Hencky strain in a fully three-dimensional setting can be found in works of its namesake. In (Hencky, 1928, Eq. (4)), we read*

$$\sigma = 2\mu\log \mathbf{V} + \lambda\operatorname{tr}(\log \mathbf{V})\mathbb{1}, \tag{2.17}$$

*which coincidentally satisfies TSTS-M$^{++}$, but cannot be derived from a strain-energy function, cf. (Yavari and Goriely, 2025, Sect. 5.4.8). To account for the latter, (Hencky, 1929, Eq. (4c)) introduced*

$$W(\mathbf{F}) = \mu\|\log \mathbf{V}\|^2 + \frac{\lambda}{2}(\operatorname{tr}\log \mathbf{V})^2 = \mu\|\log \mathbf{V}\|^2 + \frac{\lambda}{2}\log^2(\det \mathbf{F}) \tag{2.18}$$

[6] The symbol of the first Lamé constant '$\lambda$' is not to be confused with the principal stretches. Its usage should be clear from the context.

*leading to*

$$\boldsymbol{\tau} = 2\mu \log \mathbf{V} + \lambda \operatorname{tr}(\log \mathbf{V})\mathbb{1} \qquad \textit{and} \qquad \boldsymbol{\sigma} = \frac{1}{J}\left(2\mu \log \mathbf{V} + \lambda \operatorname{tr}(\log \mathbf{V})\mathbb{1}\right), \tag{2.19}$$

*which is now hyperelastic and satisfies Hill's inequality (1.7), but no longer TSTS-M$^{++}$. Interestingly, Hencky's strain-energy function (2.18) has a purely geometric interpretation in the context of geodesic distances on* $\mathrm{GL}^+(3)$*, cf. (Neff et al., 2016a, 2017).*

*More general early usage of the Hencky strain in hyperelastic modeling can be found in (Murnaghan, 1941) and in the works by (Richter, 1948, 1949). Especially noteworthy is that Richter already remarks upon the additivity of the Hencky strain for coaxial deformation states and the decomposition into deviatoric and volumetric contributions in the late 1940s, the latter of which is nowadays usually attributed to (Flory, 1961, Eq. (9)), cf. (Graban et al., 2019; Neff et al., 2020).*

*As a final comment, the lack of TSTS-M$^{++}$ in Hencky's strain-energy function (2.18) can be remedied through convexification by virtue of the exponential function such that*

$$W(\mathbf{F}) = \frac{\mu}{\alpha}\exp\left(\alpha\|\log \mathbf{V}\|^2\right) + \frac{\lambda}{2\beta}\exp\left(\beta \log^2(\det \mathbf{F})\right) + const. \quad \forall \alpha > \frac{3}{8} \quad \forall \beta > \frac{1}{8}, \tag{2.20}$$

*which then satisfies TSTS-M$^{++}$, but does not globally ensure the Legendre-Hadamard condition (1.4), cf. (Neff et al., 2015b, Sect. 4.1).*

## 3. Polyconvexity

Although the representation of necessary and sufficient conditions for polyconvexity by (Wiedemann and Peter, 2025) in terms of signed singular values constitutes a powerful tool for the construction of isotropic strain-energy functions, the omnipresent requirement of $\Pi(3)$-invariance makes a bottom-up approach by hand rather difficult. While applications such as (Neumeier et al., 2024) and (Geuken et al., 2025) work well in a computational context, analytical traceability is quickly lost. Here, a potentially less powerful representation of constitutive inequalities in terms of invariants can be beneficial. Although polyconvexity is defined as convexity with respect to $\mathbf{F}$, $\operatorname{Cof}\mathbf{F}$, and $\det\mathbf{F}$, we have to keep in mind that any strain-energy function is only implicitly parametrized by $\mathbf{F}$ through some relation that ensures objectivity and material symmetry, e.g., $W(\mathbf{F}) = \Psi(K_1, K_2, K_3)$ or $W(\mathbf{F}) = \psi(\lambda_1, \lambda_2, \lambda_3)$.

In this section, we want to present sufficient conditions for polyconvexity for a strain-energy function defined through $K_i$ given in Theorem 3.1. We will present two proofs: (i) a short one relying on the results of (Ball, 1976, Theo. 5.2); (ii) an alternative one that makes use of the norm properties of $K_i$. Notably, the usage of these invariants for the purposes of convexity are not new, e.g., cf. (Renardy, 1985, Lem. 2.1) or (Ciarlet, 1988, p. 182). Nonetheless, to the knowledge of the authors, the conditions in Theorem 3.1 have not yet been published in a comprehensive manner elsewhere, although they have much in common with (Steigmann, 2003). They also generalize some of the results by (Schröder and Neff, 2003; Hartmann and Neff, 2003) as demonstrated in Corollary 3.1.1. Nonetheless, they are by no means necessary which is straightforward to show with the help of a counter-example in Corollary 3.2.1. The sufficient conditions for polyconvexity in terms of $K_i$ are also repeated in Table 1 for quick reference.

**Theorem 3.1.** *Let*

$$W(\mathbf{F}) = \Psi(K_1, K_2, K_3), \tag{3.1}$$

*where $K_i$ are the square roots of the principal invariants of* $\mathbf{B}$*, respectively associated with* $\mathbf{F}$*,* $\operatorname{Cof}\mathbf{F}$*, and* $\det\mathbf{F}$*. If the function $\Psi$ is convex in its three arguments and non-decreasing in $K_1$ and $K_2$, then $W$ is polyconvex.*

**Proof.** With (2.2), we define

$$g(\lambda_1, \lambda_2, \lambda_3, a_1, a_2, a_3, \delta) = \Psi(K_1, K_2, K_3), \tag{3.2}$$

where

$$K_1 = \|\mathbf{F}\| = \sqrt{\lambda_1^2 + \lambda_2^2 + \lambda_3^2}, \qquad K_2 = \|\operatorname{Cof}\mathbf{F}\| = \sqrt{a_1^2 + a_2^2 + a_3^2}, \qquad \text{and} \qquad K_3 = \det\mathbf{F} = \delta \tag{3.3}$$

with $a_1 = \lambda_2\lambda_3$, $a_2 = \lambda_3\lambda_1$, and $a_3 = \lambda_1\lambda_2$.

Notice that

(i) the function $g$ remains invariant under permutation of its first three arguments due to the symmetry of $K_1$; analogous for permutations of the fourth to sixth argument due to $K_2$.
(ii) the function $g$ is non-decreasing in its first six arguments if $\Psi$ is non-decreasing in $K_1$ and $K_2$ since $K_1$ and $K_2$ are non-decreasing in $\lambda_i$ and $a_i$, respectively.
(iii) the function $g$ is convex if $\Psi$ is convex and non-decreasing in $K_1$ and $K_2$ since $K_1$ and $K_2$ are convex in $\lambda_i$ and $a_i$, respectively.

It then follows immediately from (Ball, 1976, Theo. 5.2) that $g$ and in turn $W$ is polyconvex. □

**Proof.** We define

$$\mathcal{P}(\mathbf{F}, \operatorname{Cof}\mathbf{F}, \det\mathbf{F}) = \Psi(K_1, K_2, K_3), \tag{3.4}$$

where $K_i$ are associated with $\mathbf{F}$, $\operatorname{Cof}\mathbf{F}$, and $\det\mathbf{F}$ as defined in (2.2). Note that $\mathcal{P}(\mathbf{F}, \mathbf{G}, \delta)$ takes in matrix arguments which do not have to correspond to a physical deformation state, i.e., $\mathcal{P}: \mathbb{R}^{3x3} \times \mathbb{R}^{3x3} \times \mathbb{R}^+ \rightarrow \mathbb{R}$, cf. (Ciarlet, 1988, Sect. 4.9). The definition (3.4) remains nonetheless valid since the Frobenius norm is defined for all matrices. Clearly,

$$W(\mathbf{F}) = \mathcal{P}(\mathbf{F}, \operatorname{Cof}\mathbf{F}, \det\mathbf{F}) \quad \forall\, \mathbf{F} \in \mathrm{GL}^+(3). \tag{3.5}$$

To proof that $W$ is polyconvex, we must show that $\mathcal{P}$ is convex, cf. (Ball, 1977, Theo. 2.4). Since the Frobenius norm obeys the triangle inequality and is positively homogeneous of degree one, we have

$$\|t\,\mathbf{F} + (1-t)\overline{\mathbf{F}}\| \le \|t\,\mathbf{F}\| + \|(1-t)\overline{\mathbf{F}}\| = t\|\mathbf{F}\| + (1-t)\|\overline{\mathbf{F}}\| \quad \forall\,\mathbf{F},\overline{\mathbf{F}} \in \mathbb{R}^{3x3}. \tag{3.6}$$

Thus, if the function $\Psi$ is non-decreasing in $K_1$ and $K_2$, we have

$$\begin{aligned}\mathcal{P}\big(t\,\mathbf{F} + (1-t)\overline{\mathbf{F}}, t\,\mathbf{G} + (1-t)\overline{\mathbf{G}}, t\,\delta + (1-t)\,\overline{\delta}\big) &= \Psi\big(\|t\,\mathbf{F} + (1-t)\overline{\mathbf{F}}\|, \|t\,\mathbf{G} + (1-t)\overline{\mathbf{G}}\|, t\,\delta + (1-t)\,\overline{\delta}\big)\\ &\le \Psi\big(t\|\mathbf{F}\| + (1-t)\|\overline{\mathbf{F}}\|, \|t\,\mathbf{G} + (1-t)\overline{\mathbf{G}}\|, t\,\delta + (1-t)\,\overline{\delta}\big)\\ &\le \Psi\big(t\|\mathbf{F}\| + (1-t)\|\overline{\mathbf{F}}\|, t\|\mathbf{G}\| + (1-t)\|\overline{\mathbf{G}}\|, t\,\delta + (1-t)\,\overline{\delta}\big),\end{aligned} \tag{3.7}$$

where $\mathbf{F},\overline{\mathbf{F}},\mathbf{G},\overline{\mathbf{G}} \in \mathbb{R}^{3\times3}$ and $\delta,\overline{\delta} \in \mathbb{R}^+$. Furthermore, if the function $\Psi$ is also convex in its arguments, we can continue such that

$$\begin{aligned}\mathcal{P}\big(t\,\mathbf{F} + (1-t)\overline{\mathbf{F}}, t\,\mathbf{G} + (1-t)\overline{\mathbf{G}}, t\,\delta + (1-t)\,\overline{\delta}\big) &\le \Psi\big(t\|\mathbf{F}\| + (1-t)\|\overline{\mathbf{F}}\|, t\|\mathbf{G}\| + (1-t)\|\overline{\mathbf{G}}\|, t\,\delta + (1-t)\,\overline{\delta}\big)\\ &\le t\,\Psi(\|\mathbf{F}\|, \|\mathbf{G}\|, \delta) + (1-t)\Psi(\|\overline{\mathbf{F}}\|, \|\overline{\mathbf{G}}\|, \overline{\delta})\\ &= t\,\mathcal{P}(\mathbf{F},\mathbf{G},\delta) + (1-t)\mathcal{P}(\overline{\mathbf{F}},\overline{\mathbf{G}},\overline{\delta}),\end{aligned} \tag{3.8}$$

i.e., $\mathcal{P}$ is convex and consequently $W$ is polyconvex. □

**Corollary 3.1.1.** *The functions $I_1^\alpha$ and $I_2^\alpha$ are polyconvex for $\alpha \ge \frac{1}{2}$.*

**Proof.** We take

$$W(\mathbf{F}) = \|\mathbf{F}\|^{2\alpha} \quad\Longrightarrow\quad \Psi(K_1,K_2,K_3) = K_1^{2\alpha}. \tag{3.9}$$

The results follows immediately from theorem 3.1, we require

$$\frac{\partial\Psi}{\partial K_1} = 2\alpha\,K_1^{2\alpha-1} \ge 0 \quad\text{and}\quad \frac{\partial^2\Psi}{\partial K_1^2} = 2\alpha(2\alpha-1)K_1^{2(\alpha-1)} \ge 0 \quad\Longrightarrow\quad \alpha \ge \frac{1}{2}. \tag{3.10}$$

The proof for $I_2^\alpha$ follows analogously. □

**Remark 3.2.** *In (Schröder and Neff, 2003, Proof (1)), it is shown that $I_1^\alpha$ and $I_2^\alpha$ are polyconvex for $\alpha \ge 1$. One might not expect the more general result to matter qualitatively, but as we will see in Sect. 5.2.1 it is precisely $\alpha = \frac{1}{2}$, where we find surprising material behavior. Furthermore, an input convex (partially non-decreasing) neural network defined in $K_i$ has consequently higher approximative power than one defined in $I_1$, $I_2$, and $J$, cf. (Klein et al., 2022, Rem. A.10), (Linka and Kuhl, 2023, pp. 6–7), or (Linden et al., 2023, Rem. 3.1). This extends to approaches that use an isochoric-volumetric split, cf. (Kissas et al., 2024, p. 11) or (Klein et al., 2026, Rem. 2.1).*

**Corollary 3.2.1.** *The strain-energy function*

$$W(\mathbf{F}) = \|\mathbf{F}\,\mathbf{F}^{\mathrm{T}}\|^2 - 4\det\mathbf{F} + \text{const.} \tag{3.11}$$

*is polyconvex, but does not satisfy the sufficient conditions defined in Theorem 3.1.*

**Proof.** Since

$$W(\mathbf{F}) = \|\mathbf{F}\,\mathbf{F}^{\mathrm{T}}\|^2 - 4\det\mathbf{F} \quad\Longrightarrow\quad g(\lambda_1,\lambda_2,\lambda_3,a_1,a_2,a_3,\delta) = \lambda_1^4 + \lambda_2^4 + \lambda_3^4 - 4\delta \tag{3.12}$$

it follows immediately from (Ball, 1976, Theo. 5.2) that $W$ is polyconvex.

For another more direct proof for the first term, observe that

$$\big\langle \mathrm{D}_{\mathbf{F}}\big(\|\mathbf{F}\,\mathbf{F}^{\mathrm{T}}\|^2\big), \mathbf{H}\big\rangle = 2\langle \mathbf{F}\,\mathbf{F}^{\mathrm{T}}, \mathbf{F}\,\mathbf{H}^{\mathrm{T}} + \mathbf{H}\,\mathbf{F}^{\mathrm{T}}\rangle, \tag{3.13}$$

$$\begin{aligned}\big\langle \mathrm{D}^2_{\mathbf{F}}\big(\|\mathbf{F}\,\mathbf{F}^{\mathrm{T}}\|^2\big).\mathbf{H}, \mathbf{H}\big\rangle &= 2\langle \mathbf{F}\,\mathbf{H}^{\mathrm{T}} + \mathbf{H}\,\mathbf{F}^{\mathrm{T}}, \mathbf{F}\,\mathbf{H}^{\mathrm{T}} + \mathbf{H}\,\mathbf{F}^{\mathrm{T}}\rangle + 2\langle \mathbf{F}\,\mathbf{F}^{\mathrm{T}}, \mathbf{H}\,\mathbf{H}^{\mathrm{T}} + \mathbf{H}\,\mathbf{H}^{\mathrm{T}}\rangle\\ &= 2\|\mathbf{F}\,\mathbf{H}^{\mathrm{T}} + \mathbf{H}\,\mathbf{F}^{\mathrm{T}}\|^2 + 4\langle \mathbf{F}\,\mathbf{F}^{\mathrm{T}}, \mathbf{H}\,\mathbf{H}^{\mathrm{T}}\rangle > 0,\end{aligned} \tag{3.14}$$

i.e, $\|\mathbf{F}\,\mathbf{F}^{\mathrm{T}}\|^2$ is strictly convex in $\mathbf{F}$.

From (2.1), we have $\operatorname{tr}\mathbf{B}^2 = I_1^2 - 2I_2$ and consequently

$$W(\mathbf{F}) = \|\mathbf{F}\,\mathbf{F}^{\mathrm{T}}\|^2 - 4\det\mathbf{F} \quad\Longrightarrow\quad \Psi(K_1,K_2,K_3) = K_1^4 - 2K_2^2 - 4K_3, \tag{3.15}$$

which is neither non-decreasing in $K_2$ nor convex. □

## 4. Sufficient, invariant-based conditions for TSTS-M$^{++}$

Necessary and sufficient conditions for TSTS-M$^{++}$ in terms of principal stretches are given by (Leblond, 1992, Eq. (23)) in case of hyperelasticity with

$$\mathrm{D}_{\log\mathbf{V}}\,\widehat{\sigma}(\log\mathbf{V}) \text{ is positive definite} \quad\Longleftrightarrow\quad \mathrm{D}_{\log\lambda_i}\,\widehat{\sigma}_j(\log\mathbf{V}) \text{ is positive definite}, \tag{4.1}$$

where $\widehat{\sigma}_i$ denotes the function returning the $i$th principal stress. Here, we again run into the issue that the underlying permutation invariance of $\psi(\lambda_1,\lambda_2,\lambda_3)$ seriously hinders the construction of an appropriate strain-energy function by hand. Therefore, we aim to derive a set of sufficient conditions in $K_i$ that ensure TSTS-M$^{++}$. To the knowledge of the authors, such an invariant-based result is not yet available in the literature.

From (2.5) and (2.6), we have an explicit expression connecting the derivatives of $\Psi$ and $K_i$ to the Hencky strain $\log \mathbf{V}$. It seems therefore reasonable to attempt to derive the fourth-order tensor $\mathrm{D}_{\log \mathbf{V}} \widehat{\sigma}(\log \mathbf{V})$ in closed form and to search for conditions that render it positive definite implying TSTS-M$^{++}$. This approach leads to Theorem 4.4. Before we get there, we establish two lemmas related to the definiteness of fourth-order tensors that show up in the subsequent derivation. Although the resulting sufficient conditions for TSTS-M$^{++}$ have a rather simple structure, it turns out they are not trivial to satisfy. An illustrative example for this difficulty is demonstrated in Corollary 3.1.1 for a product of monomials in $K_i$. The sufficient conditions for TSTS-M$^{++}$ in terms of $K_i$ can also be found in Table 1 for comparison with the sufficient conditions for polyconvexity.

**Lemma 4.1.** *Let* $\mathbf{B} \in \mathrm{Sym}^{++}(3)$ *and* $\mathbf{H} \in \mathrm{Sym}(3) \setminus \{\mathbf{0}\}$, *then*

$$\left\langle \left( \mathrm{D}_{\log \mathbf{V}} \mathbf{B} - 2\frac{\mathbf{B}}{K_1} \otimes \frac{\mathbf{B}}{K_1} \right) . \mathbf{H}, \mathbf{H} \right\rangle \geq 0. \tag{4.2}$$

*The inequality is strict, unless* $\mathbf{H} = H\mathbb{1}$.

**Proof.** Using the spectral decomposition (2.3), we have

$$\begin{aligned} \mathrm{D}_{\log \mathbf{V}} \mathbf{B} &= \mathrm{D}_{\log \mathbf{V}} \left( \sum_{i=1}^{3} \exp(2\log \lambda_i)\, \boldsymbol{v}_i \otimes \boldsymbol{v}_i \right) \\ &= \sum_{i=1}^{3} \mathrm{D}_{\log \mathbf{V}} \big(\exp(2\log \lambda_i)\big) \boldsymbol{v}_i \otimes \boldsymbol{v}_i + \sum_{i=1}^{3} \exp(2\log \lambda_i)\, \mathrm{D}_{\log \mathbf{V}} \big(\boldsymbol{v}_i \otimes \boldsymbol{v}_i\big) \\ &= 2\sum_{i=1}^{3} \lambda_i^2\, \boldsymbol{v}_i \otimes \boldsymbol{v}_i \otimes \boldsymbol{v}_i \otimes \boldsymbol{v}_i + \sum_{i=1}^{3}\sum_{j<i} \frac{\lambda_i^2 - \lambda_j^2}{\log \lambda_i^2 - \log \lambda_j^2} (\boldsymbol{v}_i \otimes \boldsymbol{v}_j + \boldsymbol{v}_j \otimes \boldsymbol{v}_i) \otimes (\boldsymbol{v}_i \otimes \boldsymbol{v}_j + \boldsymbol{v}_j \otimes \boldsymbol{v}_i), \end{aligned} \tag{4.3}$$

cf. (Chadwick and Ogden, 1971, Eqs. (2.1) & (2.2)) or (Itskov, 2000, Eq. (5.13)). Hence,

$$\begin{aligned} \mathrm{D}_{\log \mathbf{V}} \mathbf{B} - 2\frac{\mathbf{B}}{K_1} \otimes \frac{\mathbf{B}}{K_1} &= \frac{2}{K_1^2} \sum_{i=1}^{3}\sum_{j=1}^{3} (K_1^2 \lambda_i^2 \delta_{ij} - \lambda_i^2 \lambda_j^2) \boldsymbol{v}_i \otimes \boldsymbol{v}_i \otimes \boldsymbol{v}_j \otimes \boldsymbol{v}_j \\ &\quad + \sum_{i=1}^{3}\sum_{j<i} \frac{\lambda_i^2 - \lambda_j^2}{\log \lambda_i^2 - \log \lambda_j^2} (\boldsymbol{v}_i \otimes \boldsymbol{v}_j + \boldsymbol{v}_j \otimes \boldsymbol{v}_i) \otimes (\boldsymbol{v}_i \otimes \boldsymbol{v}_j + \boldsymbol{v}_j \otimes \boldsymbol{v}_i), \end{aligned} \tag{4.4}$$

where $\delta_{ij}$ denotes the Kronecker delta. Without loss of generality, we take $\mathbf{H} = H_{ij} \boldsymbol{v}_i \otimes \boldsymbol{v}_j = H_{ji} \boldsymbol{v}_i \otimes \boldsymbol{v}_j$ such that

$$\left\langle \left( \mathrm{D}_{\log \mathbf{V}} \mathbf{B} - 2\frac{\mathbf{B}}{K_1} \otimes \frac{\mathbf{B}}{K_1} \right) . \mathbf{H}, \mathbf{H} \right\rangle = \frac{2}{K_1^2} \sum_{i=1}^{3}\sum_{j=1}^{3} (K_1^2 \lambda_i^2 \delta_{ij} - \lambda_i^2 \lambda_j^2) H_{ii} H_{jj} + 4 \sum_{i=1}^{3}\sum_{j<i} \frac{\lambda_i^2 - \lambda_j^2}{\log \lambda_i^2 - \log \lambda_j^2} H_{ij}^2. \tag{4.5}$$

Due to the strict montonocity of the logarithm, the second term is positive, unless $H_{ij} = 0\, \forall i \neq j$. Taking a closer look at the first term, we have

$$\sum_{i=1}^{3}\sum_{j=1}^{3} \big( (\lambda_1^2 + \lambda_2^2 + \lambda_3^2) \lambda_i^2 \delta_{ij} - \lambda_i^2 \lambda_j^2 \big) H_{ii} H_{jj} = \left\langle \begin{bmatrix} H_{11} \\ H_{22} \\ H_{33} \end{bmatrix}, \begin{bmatrix} \lambda_1^2(\lambda_2^2 + \lambda_3^2) & -\lambda_1^2 \lambda_2^2 & -\lambda_1^2 \lambda_3^2 \\ -\lambda_1^2 \lambda_2^2 & \lambda_2^2(\lambda_1^2 + \lambda_3^2) & -\lambda_2^2 \lambda_3^2 \\ -\lambda_1^2 \lambda_3^2 & -\lambda_2^2 \lambda_3^2 & \lambda_3^2(\lambda_1^2 + \lambda_2^2) \end{bmatrix} \begin{bmatrix} H_{11} \\ H_{22} \\ H_{33} \end{bmatrix} \right\rangle. \tag{4.6}$$

The first and second principal invariant of this matrix are equivalent to $2I_2$ and $3I_1I_3$, respectively, while its determinant turns out to be zero. Hence, the matrix has one vanishing eigenvalue and two positive eigenvalues. The eigenvector associated with the former corresponds to $H_{11} = H_{22} = H_{33} = H$. □

**Remark 4.2.** *In case of repeating principal stretches, one encounters limiting cases in the expression for the fourth-order tensor, namely*

$$\lim_{\lambda_i \to \lambda_j} \frac{\lambda_i^2 - \lambda_j^2}{\log \lambda_i^2 - \log \lambda_j^2} = \lim_{\varepsilon \to 0} \frac{\varepsilon}{\log(\lambda_j^2 + \varepsilon) - \log \lambda_j^2} = \lambda_j^2. \tag{4.7}$$

*Additionally, the principal directions no longer correspond uniquely to one orthonormal coordinate system. In this case, we can treat* $(\boldsymbol{v}_i)_{i=1}^3$ *simply as one unspecified instance of such a principal system and the proof remains unaffected, cf. (Chadwick and Ogden, 1971, Sect. 2b).*

*Alternatively, we could have used projection tensors in the spectral decomposition (2.3) to efficiently account for repeating principal stretches, cf. (Carlson and Hoger, 1986). The corresponding representation of the fourth-order tensor derivatives unfortunately loses most of this notational simplicity as distinct expressions are needed in case of repeating principal stretches, cf. (Itskov, 2000, Eqs. (5.15) & (5.16)). The advantage of such a representation would be that the ambiguity of the eigenspace is explicitly expressed in terms of the projection tensors. This has no bearing on the proof, though.*

**Lemma 4.3.** *Let* $\mathbf{B} \in \mathrm{Sym}^{++}(3)$ *and* $\mathbf{H} \in \mathrm{Sym}(3) \setminus \{\mathbf{0}\}$, *then*

$$\left\langle \left( \mathrm{D}_{\log \mathbf{V}} \mathbf{B}^{-1} + 2K_3^2 \frac{\mathbf{B}^{-1}}{K_2} \otimes \frac{\mathbf{B}^{-1}}{K_2} \right) . \mathbf{H}, \mathbf{H} \right\rangle \leq 0. \tag{4.8}$$

*The inequality is strict, unless* $\mathbf{H} = H\mathbb{1}$.

**Proof.** Analogous to (4.4), we have

$$\mathrm{D}_{\log \mathbf{V}}\mathbf{B}^{-1} + 2K_3^2\frac{\mathbf{B}^{-1}}{K_2}\otimes\frac{\mathbf{B}^{-1}}{K_2} = -\frac{2K_3^2}{K_2^2}\sum_{i=1}^{3}\sum_{j=1}^{3}\left(\frac{K_2^2\lambda_i^{-2}\delta_{ij}}{K_3^2} - \lambda_i^{-2}\lambda_j^{-2}\right)\boldsymbol{v}_i\otimes\boldsymbol{v}_i\otimes\boldsymbol{v}_j\otimes\boldsymbol{v}_j - \sum_{i=1}^{3}\sum_{j<i}\frac{\lambda_i^{-2}-\lambda_j^{-2}}{\log\lambda_i^{-2}-\log\lambda_j^{-2}}(\boldsymbol{v}_i\otimes\boldsymbol{v}_j+\boldsymbol{v}_j\otimes\boldsymbol{v}_i)\otimes(\boldsymbol{v}_i\otimes\boldsymbol{v}_j+\boldsymbol{v}_j\otimes\boldsymbol{v}_i). \tag{4.9}$$

Notice that

$$\frac{K_2^2}{K_3^2} = \sum_{i=1}^{3}\lambda_i^{-2}, \tag{4.10}$$

i.e., expression (4.9) is equivalent to (4.4) under relabeling $\lambda_i \to \lambda_i^{-1}$. Hence, the proof of Lemma 4.1 translates directly to the current desired result, albeit with a change of sign. □

**Theorem 4.4.** *Suppose $\Psi(K_1, K_2, K_3)$ is twice continuously differentiable. The resulting elastic response function satisfies TSTS-M$^{++}$ if*

$$\Psi_1 > 0 \quad \textit{and} \quad \Psi_2 \geq 0 \quad \textit{or} \quad \Psi_1 \geq 0 \quad \textit{and} \quad \Psi_2 > 0 \tag{4.11}$$

*and*

$$\begin{bmatrix} K_1^2\Psi_{11}+K_1\Psi_1 & K_1K_2\Psi_{12} & K_1K_3\Psi_{13}-\frac{1}{2}K_1\Psi_1 \\ & K_2^2\Psi_{22}+K_2\Psi_2 & K_2K_3\Psi_{23}-\frac{1}{2}K_2\Psi_2 \\ \text{sym.} & & K_3^2\Psi_{33}\end{bmatrix} \in \mathrm{Sym}^+(3), \tag{4.12}$$

*where $\Psi_i = \frac{\partial\Psi}{\partial K_i}$ and $\Psi_{ij} = \frac{\partial^2\Psi}{\partial K_i\partial K_j}$, as well as*

$$\left\langle \begin{bmatrix}1\\2\\3\end{bmatrix}, \begin{bmatrix} K_1^2\Psi_{11}+K_1\Psi_1 & K_1K_2\Psi_{12} & K_1K_3\Psi_{13}-\frac{1}{2}K_1\Psi_1 \\ & K_2^2\Psi_{22}+K_2\Psi_2 & K_2K_3\Psi_{23}-\frac{1}{2}K_2\Psi_2 \\ \text{sym.} & & K_3^2\Psi_{33}\end{bmatrix}\begin{bmatrix}1\\2\\3\end{bmatrix}\right\rangle > 0. \tag{4.13}$$

**Proof.** Taking the tensor derivative of (2.5) with respect to $\log \mathbf{V}$ and using (2.7)–(2.9) leads to

$$\begin{aligned}\mathrm{D}_{\log\mathbf{V}}\widehat{\sigma}(\log\mathbf{V}) &= \mathrm{D}_{\log\mathbf{V}}\left(\frac{1}{K_3}\sum_{i=1}^{3}\frac{\partial\Psi}{\partial K_i}\mathrm{D}_{\log\mathbf{V}}K_i\right) \\ &= -\frac{1}{K_3^2}\sum_{i=1}^{3}\frac{\partial\Psi}{\partial K_i}\mathrm{D}_{\log\mathbf{V}}K_i\otimes\mathrm{D}_{\log\mathbf{V}}K_3 + \frac{1}{K_3}\sum_{i=1}^{3}\sum_{j=1}^{3}\frac{\partial^2\Psi}{\partial K_i\partial K_j}\mathrm{D}_{\log\mathbf{V}}K_i\otimes\mathrm{D}_{\log\mathbf{V}}K_j + \frac{1}{K_3}\sum_{i=1}^{3}\frac{\partial\Psi}{\partial K_i}\mathrm{D}^2_{\log\mathbf{V}}K_i.\end{aligned} \tag{4.14}$$

Taking a closer look at the third term, we have

$$\mathrm{D}^2_{\log\mathbf{V}}K_1 = \mathrm{D}_{\log\mathbf{V}}\left(\frac{\mathbf{B}}{K_1}\right) = \frac{1}{K_1}\left(\mathrm{D}_{\log\mathbf{V}}\mathbf{B} - \frac{\mathbf{B}}{K_1}\otimes\frac{\mathbf{B}}{K_1}\right), \tag{4.15}$$

$$\begin{aligned}\mathrm{D}^2_{\log\mathbf{V}}K_2 &= \mathrm{D}_{\log\mathbf{V}}\left(K_2\mathbb{1} - K_2^{-1}\operatorname{Cof}\mathbf{B}\right) \\ &= \left(\mathbb{1} + K_2^{-2}\operatorname{Cof}\mathbf{B}\right)\otimes\left(K_2\mathbb{1} - K_2^{-1}\operatorname{Cof}\mathbf{B}\right) - 2K_2^{-1}\operatorname{Cof}\mathbf{B}\otimes\mathbb{1} - K_2^{-1}K_3^{-2}\mathrm{D}_{\log\mathbf{V}}\mathbf{B}^{-1} \\ &= \frac{1}{K_2}\left(K_2\mathbb{1} - K_2^{-1}\operatorname{Cof}\mathbf{B}\right)\otimes\left(K_2\mathbb{1} - K_2^{-1}\operatorname{Cof}\mathbf{B}\right) - \frac{K_3^2}{K_2}\left(\mathrm{D}_{\log\mathbf{V}}\mathbf{B}^{-1} + 2K_3^2\frac{\mathbf{B}^{-1}}{K_2}\otimes\frac{\mathbf{B}^{-1}}{K_2}\right),\end{aligned} \tag{4.16}$$

$$\mathrm{D}^2_{\log\mathbf{V}}K_3 = \mathrm{D}_{\log\mathbf{V}}\left(K_3\mathbb{1}\right) = K_3\mathbb{1}\otimes\mathbb{1}. \tag{4.17}$$

Multiplying $\mathbf{H} \in \mathrm{Sym}(3)\setminus\{\mathbf{0}\}$ to both sides of (4.14) and introducing

$$x_1 = \frac{\langle\mathbf{B},\mathbf{H}\rangle}{K_1^2}, \qquad x_2 = \frac{\langle K_2^2\mathbb{1} - \operatorname{Cof}\mathbf{B},\mathbf{H}\rangle}{K_2^2}, \qquad \text{and} \qquad x_3 = \operatorname{tr}\mathbf{H}, \tag{4.18}$$

we arrive at the following inequality

$$\begin{aligned}\left\langle\mathrm{D}_{\log\mathbf{V}}\widehat{\sigma}(\log\mathbf{V}).\mathbf{H},\mathbf{H}\right\rangle &= \Psi_1\frac{1}{K_1K_3}\left\langle\left(\mathrm{D}_{\log\mathbf{V}}\mathbf{B} - 2\frac{\mathbf{B}}{K_1}\otimes\frac{\mathbf{B}}{K_1}\right).\mathbf{H},\mathbf{H}\right\rangle - \Psi_2\frac{K_3}{K_2}\left\langle\left(\mathrm{D}_{\log\mathbf{V}}\mathbf{B}^{-1} + 2K_3^2\frac{\mathbf{B}^{-1}}{K_2}\otimes\frac{\mathbf{B}^{-1}}{K_2}\right).\mathbf{H},\mathbf{H}\right\rangle \\ &\quad + \frac{1}{K_3}\left\langle\begin{bmatrix}x_1\\x_2\\x_3\end{bmatrix}, \begin{bmatrix} K_1^2\Psi_{11}+K_1\Psi_1 & K_1K_2\Psi_{12} & K_1K_3\Psi_{13}-\frac{1}{2}K_1\Psi_1 \\ & K_2^2\Psi_{22}+K_2\Psi_2 & K_2K_3\Psi_{23}-\frac{1}{2}K_2\Psi_2 \\ \text{sym.} & & K_3^2\Psi_{33}\end{bmatrix}\begin{bmatrix}x_1\\x_2\\x_3\end{bmatrix}\right\rangle > 0.\end{aligned} \tag{4.19}$$

If we require $\Psi_1$ and $\Psi_2$ to be positive and non-negative, respectively, or *vice-versa*, then by Lemma 4.1 and 4.3 the first two terms in the inequality above are positive, unless $\mathbf{H} = H\mathbb{1}$. In this case $x_1 = H$, $x_2 = 2H$, and $x_3 = 3H$, for which we require positive definiteness of the matrix of derivatives. Otherwise, semi-definiteness suffices. □

**Table 1**
Overview of the constraints placed upon a strain-energy function $W(\mathbf{F}) = \Psi(K_1, K_2, K_3)$ parametrized in terms of the invariants $K_i$ as to ensure polyconvexity or TSTS-M$^{++}$. Here, $\Psi_i = \frac{\partial \Psi}{\partial K_i}$ and $\Psi_{ij} = \frac{\partial^2 \Psi}{\partial K_i \partial K_j}$ and $\Psi$ is assumed to be twice continuously differentiable.

| Polyconvexity | True-stress-true-strain monotonicity (TSTS-M$^{++}$) |
|---|---|
| $\Psi_1 \geq 0 \quad \text{and} \quad \Psi_2 \geq 0$ | $\Psi_1 > 0 \quad \text{and} \quad \Psi_2 \geq 0 \quad \text{or} \quad \Psi_1 \geq 0 \quad \text{and} \quad \Psi_2 > 0$ |
| $\begin{bmatrix} \Psi_{11} & \Psi_{12} & \Psi_{13} \\ & \Psi_{22} & \Psi_{23} \\ \text{sym.} & & \Psi_{33} \end{bmatrix} \in \mathrm{Sym}^+(3)$ | $\begin{bmatrix} K_1^2\Psi_{11} + K_1\Psi_1 & K_1K_2\Psi_{12} & K_1K_3\Psi_{13} - \frac{1}{2}K_1\Psi_1 \\ & K_2^2\Psi_{22} + K_2\Psi_2 & K_2K_3\Psi_{23} - \frac{1}{2}K_2\Psi_2 \\ \text{sym.} & & K_3^2\Psi_{33} \end{bmatrix} \in \mathrm{Sym}^+(3)$ |
| | $\left\langle \begin{bmatrix} 1 \\ 2 \\ 3 \end{bmatrix}, \begin{bmatrix} K_1^2\Psi_{11} + K_1\Psi_1 & K_1K_2\Psi_{12} & K_1K_3\Psi_{13} - \frac{1}{2}K_1\Psi_1 \\ & K_2^2\Psi_{22} + K_2\Psi_2 & K_2K_3\Psi_{23} - \frac{1}{2}K_2\Psi_2 \\ \text{sym.} & & K_3^2\Psi_{33} \end{bmatrix} \begin{bmatrix} 1 \\ 2 \\ 3 \end{bmatrix} \right\rangle > 0$ |

**Corollary 4.4.1.** *Let $\Psi(K_1, K_2, K_3)$ be independent of $K_2$. Then sufficient condition for TSTS-M$^{++}$ are*

$$\Psi_1 > 0 \quad \text{and} \quad \begin{bmatrix} K_1^2\Psi_{11} + K_1\Psi_1 & K_1K_3\Psi_{13} - \frac{1}{2}K_1\Psi_1 \\ \text{sym.} & K_3^2\Psi_{33} \end{bmatrix} \in \mathrm{Sym}^{++}(2). \tag{4.20}$$

*Analogously for $\Psi(K_1, K_2, K_3)$ independent of $K_1$.*

**Proof.** The result follows immediately from Theorem 4.4 by restricting the reduced matrix of derivatives to be positive definite. □

**Corollary 4.4.2.** *Let*

$$\Psi(K_1, K_2, K_3) = K_1^\alpha K_2^\beta K_3^\gamma, \tag{4.21}$$

*where $\alpha, \beta, \gamma \in \mathbb{R}$. Then $\Psi$ does not satisfy the sufficient conditions for TSTS-M$^{++}$ from Theorem 4.4 for any combination of $\alpha, \beta, \gamma$.*

**Proof.** By Silvester's criterion, we check the minor of the matrix in (4.12) associated with the derivatives in $K_1$ and $K_3$. For $\Psi$ we find that

$$\begin{aligned} (K_1^2\Psi_{11} + K_1\Psi_1)K_3^2\Psi_{33} - \left(K_1K_3\Psi_{13} - \frac{1}{2}K_1\Psi_1\right)^2 &= K_1^\alpha K_2^\beta K_3^\gamma \left( \left(\alpha(\alpha-1) + \alpha\right)\gamma(\gamma-1) - \left(\alpha\,\gamma - \frac{\alpha}{2}\right)^2 \right) \\ &= -\frac{\alpha^2}{2} K_1^\alpha K_2^\beta K_3^\gamma \geq 0 \quad \Longrightarrow \quad \alpha = 0. \end{aligned} \tag{4.22}$$

Analogously, we require $\beta = 0$. This leaves $\Psi$ to be independent of $K_1$ and $K_2$, which violates the monotonicity constraints (4.11). □

## 5. The challenge questions by (Neff et al., 2024)

### 5.1. A family of chain-limited polyconvex strain-energies fulfilling TSTS-M$^{++}$

In Challenge (i) the task is to find a compressible strain-energy function which is both polyconvex and satisfies TSTS-M$^{++}$ for all $\mathbf{F} \in \mathrm{GL}^+(3)$. Equipped with the sufficient conditions from Theorems 3.1 and 4.4, one might expect that the construction of such a function is straightforward as the required monotonicity in $K_1$ and $K_2$ is shared among both constitutive constraints. Issues arise in the reconciliation of the convexity in $K_i$ and the semi-definiteness of the matrix in Theorem 4.4. So far we have been unable to square the two sets of sufficient conditions globally. It might very well be the case that this is in fact impossible, cf. (Martin et al., 2018).

One can make progress though by restricting the set of admissible deformation states. In (Neff et al., 2024, p. 64), a candidate function is proposed which is conjectured to satisfy both TSTS-M$^{++}$ and the Legendre-Hadamard condition (1.4) for restricted volumetric deformations in planar elasticity, i.e., $\mathbf{F} \in \mathrm{GL}^+(2)$, namely

$$W(\mathbf{F}) = \begin{cases} \mu \exp\left(\|\log \mathbf{V}\|^2\right) + \frac{\lambda}{2} \tan\left(\log^2(\det \mathbf{F})\right) + \text{const.}, & \text{if } \log^2(\det \mathbf{F}) < \frac{\pi}{2}, \\ \infty, & \text{else.} \end{cases} \tag{5.1}$$

Represented as floating-point numbers, the constraint reads $0.286 < \det \mathbf{F} < 3.502$. While TSTS-M$^{++}$ of the first term is established in (Neff et al., 2015a, Prop. 4.3) and TSTS-M$^{++}$ of the second term follows from its strict convexity in $J$, the Legendre-Hadamard condition is only checked numerically for $\mu = \lambda = 1$ and the set of admissible deformations up to $\|\log \mathbf{V}\| \leq 10$.

Here, we instead present rigorous proofs for three families of polyconvex strain-energy functions that satisfy TSTS-M$^{++}$ and are limited by the average deformation of line elements, area elements, and volume elements, respectively, similar to chain-limiting models, cf. (Gent, 1996). Beforehand, we briefly show that TSTS-M$^{++}$ implies TSTS-M$^{+}$ if the set of admissible Hencky strain tensors is convex.

**Proposition 5.1.** ((Neff et al., 2015a, Rem. 4.1)). *Let the elastic response function for $\sigma$ be once continuously differentiable over a convex set $C \subseteq \mathrm{Sym}(3)$ of admissible Hencky strain tensors. Then TSTS-M$^{+}$ is satisfied if*

$$\left\langle \mathrm{D}_{\log \mathbf{V}} \hat{\sigma}(\log \mathbf{V}).\mathbf{H}, \mathbf{H} \right\rangle > 0 \quad \forall \log \mathbf{V} \in C \quad \forall \mathbf{H} \in \mathrm{Sym}(3) \setminus \{\mathbf{0}\}. \tag{5.2}$$

**Proof.** Notice that

$$\begin{aligned}\left\langle \widehat{\sigma}(\log \overline{\mathbf{V}}) - \widehat{\sigma}(\log \mathbf{V}), \log \overline{\mathbf{V}} - \log \mathbf{V}\right\rangle &= \left\langle \int_0^1 \frac{\mathrm{d}}{\mathrm{d}t}\Big(\widehat{\sigma}\big(t\log\overline{\mathbf{V}} + (1-t)\log\mathbf{V}\big)\Big)\,\mathrm{d}t, \log\overline{\mathbf{V}} - \log\mathbf{V}\right\rangle \\ &= \int_0^1 \left\langle \mathrm{D}_{\log\mathbf{V}}\widehat{\sigma}(\log\mathbf{V})\Big|_{t\log\overline{\mathbf{V}}+(1-t)\log\mathbf{V}}.(\log\overline{\mathbf{V}} - \log\mathbf{V}), \log\overline{\mathbf{V}} - \log\mathbf{V}\right\rangle \mathrm{d}t.\end{aligned} \tag{5.3}$$

Consequently, the expression can be made positive by requiring that the integrand is positive which is ensured by (5.2). Importantly, the fourth-order tensor must be positive definite for all $t\log\overline{\mathbf{V}} + (1-t)\log\mathbf{V}\ \forall t \in [0,1]\ \forall \log\mathbf{V}, \log\overline{\mathbf{V}} \in \mathcal{C}$, i.e., the domain of definition must be convex. □

**Proposition 5.2.** *Let*

$$W(\mathbf{F}) = \begin{cases} -\log\big(\beta - \|\mathbf{F}\|^\alpha\big) - \gamma\log\det\mathbf{F} + \Big(\gamma - \alpha\frac{3^{\alpha/2-1}}{\beta - 3^{\alpha/2}}\Big)\det\mathbf{F} + \text{const.}, & \text{if } \|\mathbf{F}\|^\alpha < \beta, \\ \infty, & \text{else}. \end{cases} \tag{5.4}$$

*where $\alpha \geq 1$, $\beta > 3^{\alpha/2}$, and $\gamma \geq \frac{1}{4}$. Then $W$ is polyconvex and satisfies TSTS-M$^{++}$ and TSTS-M$^{+}$ within its restricted domain of definition.*

**Proof.** With (2.2) and (5.4), we have

$$\Psi(K_1, K_2, K_3) = \begin{cases} -\log\big(\beta - K_1^\alpha\big) - \gamma\log K_3 + \Big(\gamma - \alpha\frac{3^{\alpha/2-1}}{\beta - 3^{\alpha/2}}\Big)K_3 + \text{const.}, & \text{if } K_1^\alpha < \beta, \\ \infty, & \text{else}. \end{cases} \tag{5.5}$$

From (2.5), the true-stress response for $\Psi$ reads

$$\sigma = \frac{1}{K_3}\left(\frac{\alpha K_1^{\alpha-2}}{\beta - K_1^\alpha}\mathbf{B} + \Big(\gamma(K_3 - 1) - \alpha\frac{3^{\alpha/2-1}}{\beta - 3^{\alpha/2}}K_3\Big)\mathbb{1}\right), \tag{5.6}$$

which satisfies the constraint for a stress-free initial condition (2.10).

Using Theorem 3.1 for the proof of polyconvexity, it is trivial to show that the terms associated with $K_3$ are convex. Focusing on the first term in $K_1$, we have

$$\frac{\partial\Psi}{\partial K_1} = \frac{\alpha K_1^{\alpha-1}}{\beta - K_1^\alpha} > 0 \qquad \text{and} \qquad \frac{\partial^2\Psi}{\partial K_1^2} = \alpha\left(\frac{(\alpha-1)K_1^{\alpha-2}}{\beta - K_1^\alpha} + \frac{\alpha K_1^{2(\alpha-1)}}{(\beta - K_1^\alpha)^2}\right) > 0. \tag{5.7}$$

Since the constraint on the admissible deformation states is defined in terms of a convex function in $\mathbf{F}$, the restricted domain remains a convex set for the definition of $\mathcal{P}(\mathbf{F}, \mathbf{G}, \delta)$. Consequently, $W$ is polyconvex.

The majority of the sufficient conditions for polyconvexity carry over to the ones from TSTS-M$^{++}$ in Corollary 4.4.1. Indeed, all that is left to show is that the matrix in (4.20) is positive definite by Silvester's criterion with

$$(K_1^2\Psi_{11} + K_1\Psi_1)K_3^2\Psi_{33} - \Big(K_1K_3\Psi_{13} - \frac{1}{2}K_1\Psi_1\Big)^2 > 0 \tag{5.8}$$

leading to

$$\frac{\gamma\,\alpha(\alpha-1)K_1^\alpha}{\beta - K_1^\alpha} + \frac{\gamma\,\alpha^2K_1^{2\alpha}}{(\beta - K_1^\alpha)^2} + \frac{\gamma\,\alpha K_1^\alpha}{\beta - K_1^\alpha} - \left(\frac{1}{2}\frac{\alpha K_1^\alpha}{\beta - K_1^\alpha}\right)^2 = \frac{\gamma\,\alpha^2K_1^\alpha}{\beta - K_1^\alpha} + \Big(\gamma - \frac{1}{4}\Big)\frac{\alpha^2K_1^{2\alpha}}{(\beta - K_1^\alpha)^2} > 0. \tag{5.9}$$

This completes the proof of TSTS-M$^{++}$. Consequently, $W$ automatically leads to a valid elastic law in the infinitesimal theory adhering to (2.12). For the sake of completeness, we have

$$\mu = \alpha\frac{3^{\alpha/2-1}}{\beta - 3^{\alpha/2}} > 0 \qquad \text{and} \qquad \kappa = \gamma + \frac{\alpha\,3^{\alpha/2}\big(\alpha\beta - 3(\beta - 3^{\alpha/2})\big)}{9(\beta - 3^{\alpha/2})^2} > 0 \tag{5.10}$$

with $\nu \in \big(-1, \frac{1}{2}\big)$.

For the implication of TSTS-M$^{+}$ via Proposition (5.2), we need to show that the set of admissible Hencky strain tensors is convex, i.e., that $K_1$ is convex in $\log\mathbf{V}$. This can be straightforwardly proven by expressing $K_1$ in terms of $\log\lambda_i$, i.e, $K_1 = \sqrt{\exp(2\log\lambda_1) + \exp(2\log\lambda_2) + \exp(2\log\lambda_3)}$. The expression is convex in $\log\lambda_i$ and therefore also in $\log\mathbf{V}$, cf. (Hill, 1968, p. 238). □

**Remark 5.3.** *Another more direct way to see that the first term in (5.4) is polyconvex is to differentiate by $\mathbf{F}$ such that*

$$\big\langle \mathrm{D}_{\mathbf{F}}\big(-\log(\beta - \|\mathbf{F}\|^\alpha)\big), \mathbf{H}\big\rangle = \alpha\big(\beta - \|\mathbf{F}\|^\alpha\big)^{-1}\|\mathbf{F}\|^{\alpha-2}\langle\mathbf{F}, \mathbf{H}\rangle, \tag{5.11}$$

$$\begin{aligned}\big\langle \mathrm{D}^2_{\mathbf{F}}\big(-\log(\beta - \|\mathbf{F}\|^\alpha)\big).\mathbf{H}, \mathbf{H}\big\rangle &= \big(\alpha\big(\beta - \|\mathbf{F}\|^\alpha\big)^{-1}\|\mathbf{F}\|^{\alpha-2}\langle\mathbf{F}, \mathbf{H}\rangle\big)^2 + \alpha(\alpha-1)\big(\beta - \|\mathbf{F}\|^\alpha\big)^{-1}\|\mathbf{F}\|^{\alpha-4}\langle\mathbf{F}, \mathbf{H}\rangle^2 \\ &\quad + \alpha\big(\beta - \|\mathbf{F}\|^\alpha\big)^{-1}\|\mathbf{F}\|^{\alpha-4}(\|\mathbf{F}\|^2\|\mathbf{H}\|^2 - \langle\mathbf{F}, \mathbf{H}\rangle^2) > 0.\end{aligned} \tag{5.12}$$

*The last term is non-negative by virtue of the Cauchy-Schwarz inequality.*

**Proposition 5.4.** *Let*

$$W(\mathbf{F}) = \begin{cases} -\log\left(\beta - \|\mathrm{Cof}\,\mathbf{F}\|^{\alpha}\right) - \gamma \log\det\mathbf{F} + \left(\gamma - 2\alpha\frac{3^{\alpha/2-1}}{\beta-3^{\alpha/2}}\right)\det\mathbf{F} + \text{const.}, & \text{if } \|\mathrm{Cof}\,\mathbf{F}\|^{\alpha} < \beta, \\ \infty, & \text{else}. \end{cases} \tag{5.13}$$

*where $\alpha \geq 1$, $\beta > 3^{\alpha/2}$, and $\gamma \geq \frac{1}{4}$. Then $W$ is polyconvex and satisfies TSTS-M$^{++}$ and TSTS-M$^{+}$ within its restricted domain of definition.*

**Proof.** Since the sufficient conditions from Theorems 3.1 and 4.4 are invariant under relabeling of $K_1$ and $K_2$, the majority of the proof of Proposition 5.2 carries over directly. This includes the convexity of the domain since $K_2$ is convex in both $\mathrm{Cof}\,\mathbf{F}$ and $\log\mathbf{V}$. The latter can again be proven by expressing

$$K_2 = \sqrt{\exp\left(2(\log\lambda_1 + \log\lambda_2)\right) + \exp\left(2(\log\lambda_2 + \log\lambda_3)\right) + \exp\left(2(\log\lambda_3 + \log\lambda_1)\right)}, \tag{5.14}$$

which is convex in $\log\lambda_i$.

A small adjustment must be made to the third term to ensure a stress-free initial configuration, albeit without consequences for polyconvexity and TSTS-M$^{++}$. The elastic constants of the infinitesimal theory read

$$\mu = \alpha\frac{3^{\alpha/2-1}}{\beta - 3^{\alpha/2}} > 0 \qquad \text{and} \qquad \kappa = \gamma + \frac{2\alpha 3^{\alpha/2}\left(2\alpha\beta - 3(\beta - 3^{\alpha/2})\right)}{9(\beta - 3^{\alpha/2})^2} > 0 \tag{5.15}$$

with $\nu \in \left(-1, \frac{1}{2}\right)$. □

**Proposition 5.5.** *Let*

$$W(\mathbf{F}) = \begin{cases} \frac{\|\mathbf{F}\|^3}{\beta - \log^2(\det\mathbf{F})} - \frac{3\sqrt{3}}{\beta}\det\mathbf{F} + \text{const.}, & \text{if } \log^2(\det\mathbf{F}) < \beta, \\ \infty, & \text{else}. \end{cases} \tag{5.16}$$

*where $0 < \beta < \frac{27}{4}$. Then $W$ is polyconvex and satisfies TSTS-M$^{++}$ and TSTS-M$^{+}$ within its restricted domain of definition.*

**Proof.** We rewrite (5.16) with (2.2) into

$$\Psi(K_1, K_2, K_3) = \begin{cases} \frac{K_1^3}{\beta - \log^2 K_3} - \frac{3\sqrt{3}}{\beta}K_3 + \text{const.}, & \text{if } \log^2 K_3 < \beta, \\ \infty, & \text{else}. \end{cases} \tag{5.17}$$

The true-stress response follows from (2.5) with

$$\boldsymbol{\sigma} = \frac{1}{K_3}\left(\frac{3K_1}{\beta - \log^2 K_3}\mathbf{b} + \left(\frac{2K_1^3}{(\beta - \log^2 K_3)^2}\log K_3 - \frac{3\sqrt{3}}{\beta}K_3\right)\mathbb{1}\right). \tag{5.18}$$

It is straightforward to verify that the constraint of a stress-free initial configuration (2.10) is satisfied.

For notational brevity, we introduce

$$u(K_3) = \frac{1}{\beta - \log^2 K_3} > 0. \tag{5.19}$$

With the sufficient conditions for polyconvexity from Theorem 3.1 and Silvester's criterion, we have

$$\frac{\partial\Psi}{\partial K_1} = 3K_1^2 u > 0, \qquad \frac{\partial^2\Psi}{\partial K_1^2} = 6K_1 u > 0, \tag{5.20}$$

and

$$\Psi_{11}\Psi_{33} - \Psi_{13}^2 = 3K_1^4\left(2\,u\,u'' - 3(u')^2\right) > 0, \tag{5.21}$$

where the prime denotes differentiation with respects to $K_3$. To show the last condition indeed holds we reinsert the abbreviation (5.19) to end up with

$$\begin{aligned} 2u\,u'' - 3(u')^2 &= \frac{2}{(\beta - \log^2 K_3)^3}\left(\frac{1}{\beta - \log^2 K_3}\frac{8\log^2 K_3}{K_3^2} + \frac{2(1 - \log K_3)}{K_3^2}\right) - \frac{1}{(\beta - \log^2 K_3)^4}\frac{12\log K_3^2}{K_3^2} \\ &= \frac{4}{K_3^2(\beta - \log^2 K_3)^4}\left(\log^3 K_3 - \beta\log K_3 + \beta\right). \end{aligned} \tag{5.22}$$

With $x = \log K_3$, we have the depressed cubic

$$f(x) = x^3 - \beta x + \beta, \tag{5.23}$$

for which $f(0) = \beta > 0$ and which does not cross the abscissa since it does not have any real roots as long as the discriminant $\Delta = 4b^3 - 27b^2 = \beta^2(4\beta - 27) < 0$ remains negative. Consequently, $f(x) > 0$ which establishes (5.21). Since the constraint

$$\log^2\det\mathbf{F} < \beta \qquad \Longrightarrow \qquad \exp(-\sqrt{\beta}) < \det\mathbf{F} < \exp(\sqrt{\beta}), \tag{5.24}$$

the restricted domain remains a convex set for the definition of $\mathcal{P}(\mathbf{F},\mathbf{G},\delta)$. Hence, $W$ is polyconvex. In other word, the term $\frac{\|\mathbf{F}\|^3}{\beta-\log^2(\det\mathbf{F})}$ is convex in $\mathbf{F}$ and $\det\mathbf{F}$ for all $\log^2(\det\mathbf{F}) < \beta$.

From Corollary 4.4.1, we have sufficient conditions for TSTS-M$^{++}$ which are largely already satisfied by (5.20). It remains to show that the determinant of the matrix in (4.12) is positive, i.e.,

$$(K_1^2\Psi_{11} + K_1\Psi_1)K_3^2\Psi_{33} - \left(K_1K_3\Psi_{13} - \frac{1}{2}K_1\Psi_1\right)^2 = 9K_1^6\left(K_3^2\,u\,u'' - \left(K_3\,u' - \frac{u}{2}\right)^2\right) > 0 \tag{5.25}$$

Again, reinserting (5.19), leads to

$$\begin{aligned} K_3^2\,u\,u'' - \left(K_3\,u' - \frac{u}{2}\right)^2 &= \frac{K_3^2}{(\beta-\log^2 K_3)^3}\left(\frac{1}{\beta-\log^2 K_3}\frac{8\log^2 K_3}{K_3^2} + \frac{2(1-\log K_3)}{K_3^2}\right) \\ &\quad - \left(\frac{2\log K_3}{(\beta-\log^2 K_3)^2} - \frac{1}{2(\beta-\log^2 K_3)}\right)^2 \\ &= -\frac{\log^4 K_3 - 2(\beta+4)\log^2 K_3 + \beta(\beta-8)}{4(\beta-\log^2 K_3)}. \end{aligned} \tag{5.26}$$

To show that this expression and in turn (5.25) is positive, we use a similar trick to before. Observe that the numerator again looks like a polynomial with $x = \log^2 K_3$ such that

$$f(x) = -x^2 + 2(\beta+4)x - \beta(\beta-8) \tag{5.27}$$

Remembering $\beta \in \left(0, \frac{27}{4}\right)$, it follows that $f(0) = -\beta(\beta-8) > 0$. Here, the discriminant reads $\Delta = 4(\beta+4)^2 - 4\beta(\beta-8) = 64(\beta+1) > 0$ and, given the positivity of the second and third coefficient in $f(x)$, we have one positive and one negative root. Taking the relevant former one and denoting it with $x^*$, we have

$$x^* = \log^2 K_3^* = \beta + 4 + 4\sqrt{\beta+1}. \tag{5.28}$$

Consequently, the polynomial $f(x)$ crosses the abscissa outside the set of admissible deformation states and remains positive within if

$$\beta \le \log^2 K_3^* = \beta + 4 + 4\sqrt{\beta+1} \quad \Longrightarrow \quad 1 + \sqrt{1+\beta} \ge 0, \tag{5.29}$$

which is indeed the case. Hence, (5.26) and in turn (5.25) are positive and TSTS-M$^{++}$ holds within the restricted domain.

A pleasant side effect of TSTS-M$^{++}$ is that the linearization condition (2.12) is already taken care of. We nonetheless provide the elastic constant of the infinitesimal theory reading

$$\mu = \frac{3\sqrt{3}}{\beta}, \qquad \kappa = \frac{6\sqrt{3}}{\beta^2}, \qquad \text{and} \qquad \nu = -\frac{\beta-3}{\beta+6} \in \left(-\frac{5}{17}, \frac{1}{2}\right). \tag{5.30}$$

The implication of TSTS-M$^{+}$ from Proposition 5.1 follows by noticing that the constraint

$$\log^2\det\mathbf{F} = (\operatorname{tr}\log\mathbf{V})^2 < \beta, \tag{5.31}$$

is convex in $\log\mathbf{V}$. Hence, the set of admissible Hencky strain tensor is also convex. □

**Remark 5.6.** *Another family of strain-energy functions can be acquired by swapping out $\|\mathbf{F}\|$ for $\|\operatorname{Cof}\mathbf{F}\|$ in (5.16). The whole proof remains virtually the same due to the symmetries in Theorem 3.1 and Corollary 4.4.1 regarding $K_1$ and $K_2$, analogous to Proposition 5.4. Solely the term related to the stress-free initial condition and hence the elastic constants of the infinitesimal theory must be slightly adjusted.*

### 5.2. *True-stress monotonicity in unconstrained uniaxial extension-compression*

Here, we want to give a family of solutions to Challenge (ii) and an interesting result concerning Challenge (iii). First some clarifying definitions. By unconstrained uniaxial extension-compression along $\boldsymbol{e}_1$, we refer to an irrotational, spatially homogeneous deformation

$$\mathbf{F} = \sum_{i=1}^{3} \lambda_i\,\boldsymbol{e}_i \otimes \boldsymbol{e}_i, \tag{5.32}$$

where $\lambda_1$ is prescribed, resulting in Cauchy stress tensor $\boldsymbol{\sigma} = \sigma_{11}\boldsymbol{e}_1 \otimes \boldsymbol{e}_1$. The spatial homogeneity of both the deformation and the resulting Cauchy stress tensor satisfy the local balance of linear momentum trivially. With the boundary conditions we then recover a system of equations, the solution of which implies a function for $\sigma_{11}$ given $\lambda_1$.

By implication (1.14) and (1.16), TSTS-M$^{++}$ ensures that the stress response in unconstrained uniaxial extension-compression must be strictly monotonic. On the other hand, it is not immediately obvious whether polyconvexity ensures such a global statement.

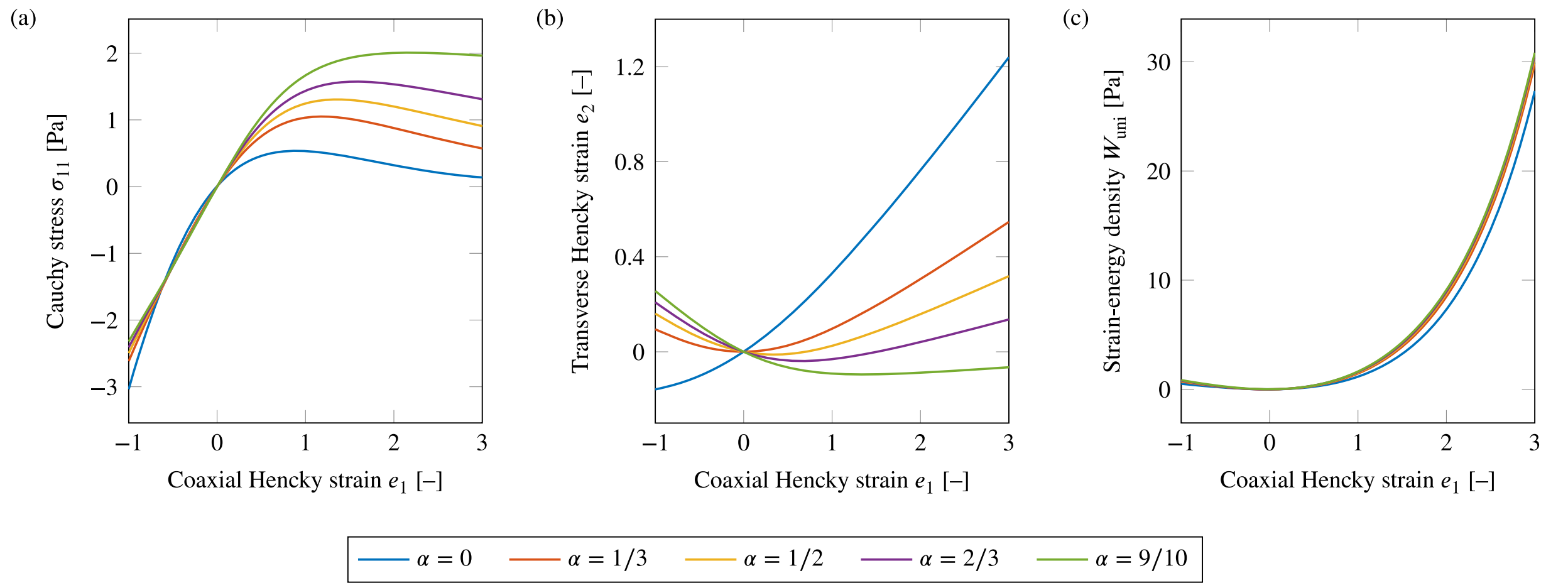

**Fig. 2.** Trajectory of (a) the non-monotonic Cauchy stress $\sigma_{11}$, (b) the transverse Hencky strain $\log \lambda_2$, and (c) the strain energy density $W_{\text{uni}}$ during uniaxial unconstrained tension-compression given the strain-energy function (5.37) for some applied coaxial Hencky strain $\log \lambda_1$. The legend applies throughout.

### *5.2.1. The compressible case*

Together with the isotropic elastic response function from (2.5), the problem statement reduces to solving the following system of equations

$$\boldsymbol{\sigma} = \sigma_{11}\, \boldsymbol{e}_1 \otimes \boldsymbol{e}_1 = \frac{1}{K_3} \sum_{i=1}^{3} \left( \frac{\partial \Psi}{\partial K_1} \frac{\lambda_i^2}{K_1} + \frac{\partial \Psi}{\partial K_2} \frac{K_2^2 - K_3^2 \lambda_i^{-2}}{K_2} + \frac{\partial \Psi}{\partial K_3} K_3 \right) \boldsymbol{e}_i \otimes \boldsymbol{e}_i, \tag{5.33}$$

where $K_i$ as in (3.3), which constitutes three equations for the three unknowns $\sigma_{11}$, $\lambda_2$, and $\lambda_3$, while the coaxial stretch $\lambda_1$ is given. The scalar equations associated with $\boldsymbol{e}_2$ and $\boldsymbol{e}_3$ are identical and we can directly reduce the system by taking $\lambda_2 = \lambda_3$. This equivalence of $\lambda_2$ and $\lambda_3$ is physically self-evident due to isotropy. We are left with

$$\frac{\partial \Psi}{\partial K_1} \frac{\lambda_2^2}{K_1} + \frac{\partial \Psi}{\partial K_2} \frac{K_2^2 - K_3^2 \lambda_2^{-2}}{K_2} + \frac{\partial \Psi}{\partial K_3} K_3 = 0, \tag{5.34}$$

which defines an implicit relation between $\lambda_1$ and $\lambda_2$. The remaining equation

$$\sigma_{11} = \frac{1}{K_3} \frac{\partial \Psi}{\partial K_1} \frac{\lambda_1^2}{K_1} + \frac{1}{K_3} \frac{\partial \Psi}{\partial K_2} \frac{K_2^2 - K_3^2 \lambda_1^{-2}}{K_2} + \frac{\partial \Psi}{\partial K_3} \tag{5.35}$$

together with the transverse-stretch relation closes the problem.

Notably, in the case of unconstrained uniaxial compression, it is shown numerically in (Korobeynikov et al., 2025, Table 2, Fig. 15c) that the polyconvex strain-energy function

$$W(\mathbf{F}) = \frac{\|\mathbf{F}\|^2}{(\det \mathbf{F})^{2/3}} + \frac{2}{3} \frac{1+\nu}{1-2\nu} (\det \mathbf{F} - 1)^2 + \text{const.} \quad \forall \nu \in \left\{ \frac{2}{5}, \frac{9}{20} \right\} \tag{5.36}$$

leads to a non-monotonic true-stress response in the sense that there exist multiple true-stress states $\sigma_{11}$ for some coaxial stretch $\lambda_1$. Here, we provide a non-monotonic example in tension, where distinct coaxial stretches $\lambda_1$ can lead to the same Cauchy stress $\sigma_{11}$.

**Proposition 5.7.** *Let*

$$W(\mathbf{F}) = \sqrt{3} \|\mathbf{F}\| + \frac{1}{\alpha\, (\det \mathbf{F})^{\alpha}} + \text{const.}, \tag{5.37}$$

*where $\alpha \in [0, 1)$. Then the elastic response function derived from the polyconvex strain-energy function $W$ leads to a proper linearization in accordance with the infinitesimal theory and shows a non-monotonic true-stress trajectory in unconstrained uniaxial extension.*

**Proof.** With (5.37), we have

$$\Psi(K_1, K_2, K_3) = \sqrt{3} K_1 + \frac{1}{\alpha} K_3^{-\alpha} + \text{const.} \tag{5.38}$$

The polyconvexity of $W$ follows directly from the sufficient condition outlined in Theorem 3.1 since

$$\frac{\partial \Psi}{\partial K_1} = \sqrt{3}, \qquad \frac{\partial^2 \Psi}{\partial K_1^2} = 0, \qquad \text{and} \qquad \frac{\partial^2 \Psi}{\partial K_3^2} = (\alpha + 1) K_3^{-(\alpha+2)} \geq 0. \tag{5.39}$$

Furthermore, the undeformed configuration is stress-free by satisfying (2.10). Using the expression from (Truesdell and Noll, 1965, Eq. (50.13)) and (2.14), we arrive at the linearized constants

$$\mu = 1, \qquad \kappa = \alpha + \frac{1}{3} > 0, \qquad \text{and} \qquad \nu = \frac{3\alpha - 1}{6\alpha + 4} \in \left[-\frac{1}{4}, \frac{1}{5}\right), \tag{5.40}$$

i.e., a proper linearly elastic law in the infinitesimal theory.

Plugging $\Psi$ into (5.34) and remembering $\lambda_2 = \lambda_3$, we read

$$\frac{\sqrt{3}}{K_1}\lambda_2^2 - K_3^{-\alpha} = 0 \quad \Longrightarrow \quad f(\lambda_1, \lambda_2) = \lambda_2^{4(1+\alpha)} - \frac{2}{3}\lambda_1^{-2\alpha}\lambda_2^2 - \frac{1}{3}\lambda_1^{2(1-\alpha)} = 0. \tag{5.41}$$

With the implicit function theorem in mind, we evaluate the partial derivative of $f$ with respect to $\lambda_2$ at a solution point and find

$$\frac{\partial f}{\partial \lambda_2} = 4(1+\alpha)\lambda_2^{3+4\alpha} - \frac{4}{3}\lambda_1^{-2\alpha}\lambda_2 = \frac{2}{\lambda_2}\left((1+2\alpha)\lambda_2^{4(1+\alpha)} + \frac{1}{3}\lambda_1^{2(1-\alpha)}\right) > 0. \tag{5.42}$$

Additionally, using a generalization of Descartes' rule of signs for real valued exponents, we can deduce that there exist only a single positive solution to $f(\lambda_1, \lambda_2) = 0$ for every $\lambda_1$, cf. (Wang, 2004). This establishes a surjective continuously differentiable function for the transverse stretch over the applied coaxial stretch, i.e., $\lambda_2 = \lambda_2(\lambda_1)$.

Notice that for all $\alpha \in [0, 1)$, we have

$$\lambda_2^2\left(\lambda_2^{2(1+2\alpha)} - \frac{2}{3}\lambda_1^{-2\alpha}\right) - \frac{1}{3}\lambda_1^{2(1-\alpha)} = 0 \quad \Longrightarrow \quad \lim_{\lambda_1 \to \infty} \lambda_2(\lambda_1) = \infty \tag{5.43}$$

Taking a look at (5.35) for $\Psi$, we have

$$\sigma_{11}(\lambda_1) = \frac{\sqrt{3}}{K_1 K_3}\lambda_1^2 - K_3^{-\alpha-1}, \tag{5.44}$$

which together with the properties of the transverse-stretch relation $\lambda_2(\lambda_1)$ implies a continuously differentiable function for $\sigma_{11}$ in $\lambda_1$. With (5.43),

$$\lim_{\lambda_1 \to \infty} (\lambda_1 \lambda_2^2)^{-\alpha-1} = 0 \qquad \text{and} \qquad \lim_{\lambda_1 \to \infty} \frac{\lambda_1^2}{K_1 K_3} = \lim_{\lambda_1 \to \infty} \left(\lambda_2^4 + 2\lambda_2^6\lambda_1^{-2}\right)^{-1/2} = 0, \tag{5.45}$$

which implies $\lim_{\lambda_1 \to \infty} \sigma_{11}(\lambda_1) = 0$. The stress $\sigma_{11}$ must also vanish in the undeformed configuration and its trajectory has a positive initial slope due to (5.40). Consequently, by virtue of Rolle's theorem, the response function for $\sigma_{11}$ has a maximum in tension and a non-monotonic trajectory. □

**Remark 5.8.** *In Fig. 2 we visualize a family of curves for the Cauchy stress $\sigma_{11}$, the transverse Hencky strain $\log\lambda_2$, and the associated strain-energy density $W_{\text{uni}}$ over the applied coaxial Hencky strain $\log\lambda_1$. Here, $W_{\text{uni}}(\log\lambda_1)$ refers to $W(\mathbf{F})$ evaluated for the uniaxial deformation (5.32) satisfying the transverse-stretch relation (5.41). Said implicit relation is solved numerically using Julia, cf. (Bezanson et al., 2017). Interestingly, $W_{\text{uni}}$ appears to be convex in $\log\lambda_1$, although we have not rigorously proven this claim. The material is only initially auxetic for $\alpha \in \left[0, \frac{1}{3}\right)$. The transverse stretch diverges however for all allowed values of $\alpha$ implying a local minimum in the transverse stretch trajectory for $\alpha \in \left(\frac{1}{3}, 1\right)$. For $\alpha = 0$ and $\alpha = \frac{1}{2}$, the implicit relation for the transverse stretch is a quadratic and depressed cubic equation, respectively, and can be solved in closed form.*

**Remark 5.9.** *Interestingly, the strain-energy function (5.37) satisfies Hill's inequality. This is straightforward to see by parametrizing $W(\mathbf{F})$ in terms of the Hencky strain $\log\mathbf{V}$, i.e.,*

$$\begin{aligned}\widehat{W}(\log\mathbf{V}) &= \sqrt{3}\|\exp\log\mathbf{V}\| + \frac{1}{\alpha}\exp(-\alpha\,\mathrm{tr}\log\mathbf{V}) + const. \\ &= \sqrt{3\big(\exp(2\log\lambda_1) + \exp(2\log\lambda_2) + \exp(2\log\lambda_3)\big)} + \frac{1}{\alpha}\exp\big(-\alpha(\log\lambda_1 + \log\lambda_2 + \log\lambda_3)\big) + const.\end{aligned} \tag{5.46}$$

*which is strictly convex in $\log\lambda_i$ and therefore also in $\log\mathbf{V}$, cf. (Hill, 1968, p. 238). The resulting non-monotonicity is therefore another example for the inadequacies of Hill's inequality as a general constitutive constraint in case of compressible material behavior.*

### 5.2.2. The incompressible case

Challenge (iii) asks for an incompressible strain-energy function that leads to a non-monotonic true-stress response in unconstrained uniaxial extension-compression. Although we are unable to provide an example, we can identify a set of necessary conditions which need to be satisfied. For this purposes, we are working with the representation of the isotropic strain-energy function in terms of principal stretches through $\psi(\lambda_1, \lambda_2, \lambda_3)$. From (2.11) and (2.15), we have

$$\sigma = -p\mathbb{1} + \mathrm{D}_{\log\mathbf{V}}\widehat{W}(\log\mathbf{V}) = \sum_{i=1}^{3}\left(-p + \lambda_i\frac{\partial\psi}{\partial\lambda_i}\right)\boldsymbol{v}_i \otimes \boldsymbol{v}_i. \tag{5.47}$$

The deformation gradient (5.32) applies here as well, albeit with $\lambda_1\lambda_2\lambda_3 = 1$ due to the incompressiblity constraint. Hence, we require

$$\sigma = \sigma_{11}\boldsymbol{e}_1 \otimes \boldsymbol{e}_1 = \sum_{i=1}^{3}\left(-p + \lambda_i\frac{\partial\psi}{\partial\lambda_i}\right)\boldsymbol{e}_i \otimes \boldsymbol{e}_i. \tag{5.48}$$

Here, the unknowns are $p$, $\lambda_2$, and $\lambda_3$. As with the compressible case, we can immediately satisfy one equation by taking $\lambda_2 = \lambda_3$ and we have $\lambda_2 = \lambda_1^{-1/2}$ from the incompressibility constraint. The Lagrange parameter $p$ also follows immediately with

$$p = \lambda_2 \frac{\partial \psi}{\partial \lambda_2} = \lambda_3 \frac{\partial \psi}{\partial \lambda_3} \quad \Longrightarrow \quad \sigma_{11} = \lambda_1 \frac{\partial \psi}{\partial \lambda_1} - \frac{\lambda_1^{-1/2}}{2}\left(\frac{\partial \psi}{\partial \lambda_2} + \frac{\partial \psi}{\partial \lambda_3}\right). \tag{5.49}$$

This representation brings us to the following – to the knowlegde of the authors – previously unknown observation.

**Proposition 5.10.** *If a continuously differentiable incompressible strain-energy function $W$ satisfies the sufficient conditions for polyconvexity proposed by (Ball, 1976, Theo. 5.2), then its true-stress response in unconstrained uniaxial extension-compression is monotonic.*

**Proof.** We abbreviate $x = \log \lambda_1$ and define

$$\phi(x) = \psi\left(\exp(x), \exp\left(-\frac{x}{2}\right), \exp\left(-\frac{x}{2}\right)\right), \tag{5.50}$$

such that $\lambda_1 = \exp(x)$ and $\lambda_2 = \lambda_3 = \exp\left(-\frac{x}{2}\right) = \lambda_1^{-1/2}$. Taking the derivative of $\phi$ with respect to $x$ and applying the chain rule, we find

$$\frac{\mathrm{d}\phi}{\mathrm{d}x} = \frac{\partial \psi}{\partial \lambda_1}\exp(x) - \frac{\partial \psi}{\partial \lambda_2}\frac{\exp\left(-\frac{x}{2}\right)}{2} - \frac{\partial \psi}{\partial \lambda_3}\frac{\exp\left(-\frac{x}{2}\right)}{2} = \lambda_1 \frac{\partial \psi}{\partial \lambda_1} - \frac{\lambda_1^{-1/2}}{2}\left(\frac{\partial \psi}{\partial \lambda_2} + \frac{\partial \psi}{\partial \lambda_3}\right), \tag{5.51}$$

which is identical to (5.49), i.e., we can derive the stress response of an incompressible hyperelastic solid in unconstrained uniaxial extension-compression from the potential $\phi$. In fact, the expression (5.51) is closely related to the Murnaghan-Richter formula (1.9).

Taking

$$\psi(\lambda_1, \lambda_2, \lambda_3) = g(\lambda_1, \lambda_2, \lambda_3, \lambda_2\lambda_3, \lambda_3\lambda_1, \lambda_1\lambda_2), \tag{5.52}$$

it follows from (5.50) that

$$\phi(x) = g\left(\exp(x), \exp\left(-\frac{x}{2}\right), \exp\left(-\frac{x}{2}\right), \exp(-x), \exp\left(\frac{x}{2}\right), \exp\left(\frac{x}{2}\right)\right). \tag{5.53}$$

If and only if $g$ is convex and non-decreasing in its arguments, then $g$ fulfills the sufficient condition for polyconvexity by (Ball, 1976, Theo. 5.2) in case of incompressibility, cf. (Ball, 1976, Sect. 8) and (Ball, 1977, Item (H1)'). Since the exponential function is also convex, it follows that $\phi$ must be convex which in turn forces a monotonic true-stress response by virtue of (5.51). □

**Remark 5.11.** *In (Rosakis, 1997, Rem. 3.1) and (Wiedemann and Peter, 2025), it is shown that the monotonicity requirement in (Ball, 1976, Theo. 5.2) is too strict. Consequently, we cannot conclude that it is impossible to have an incompressible polyconvex hyperelastic material that produces a non-monotonic true-stress response in unconstrained uniaxial extension-compression. Then again, we have not been able to come up with a polyconvex incompressible strain-energy function which violates the monotonicity constraint as construction by hand is made difficult by the $\Pi(3)$-invariance requirement, cf. (Wiedemann and Peter, 2025, Sect. 2). The search for a valid candidate could be attempted computationally by a universal function approximator, cf. (Geuken et al., 2025).*

**Remark 5.12.** *For incompressible material behavior, TSTS-M$^{++}$ reduces to Hill's inequality, the latter of which is satisfied if $W$ is convex in $\log \mathbf{V}$, i.e.*

$$\left\langle \mathrm{D}^2_{\log \mathbf{V}} \widehat{W}(\log \mathbf{V}).\mathbf{H}, \mathbf{H} \right\rangle > 0 \quad \forall \mathbf{H} \in \mathrm{Sym}(3) \setminus \{\mathbf{0}\}, \tag{5.54}$$

*since $\boldsymbol{\tau} = \mathrm{D}_{\log \mathbf{V}} \widehat{W}$ by way of (1.9).*

*Defining some incompressible strain-energy function $W(\mathbf{F}) = \Psi(K_1, K_2, K_3)$ independent of $K_3$, we have*

$$\mathrm{D}^2_{\log \mathbf{V}} \widehat{W}(\log \mathbf{V}) = \sum_{i=1}^{2}\sum_{j=1}^{2} \frac{\partial^2 \Psi}{\partial K_i \partial K_j} \mathrm{D}_{\log \mathbf{V}} K_i \otimes \mathrm{D}_{\log \mathbf{V}} K_j + \sum_{i=1}^{2} \frac{\partial \Psi}{\partial K_i} \mathrm{D}^2_{\log \mathbf{V}} K_i. \tag{5.55}$$

*Following the approach used in the proof of Theorem 4.4, we arrive at sufficient conditions for (5.54) in $K_i$ with*

$$\Psi_1 > 0 \quad \textit{and} \quad \Psi_2 \geq 0 \quad \textit{or} \quad \Psi_1 \geq 0 \quad \textit{and} \quad \Psi_2 > 0 \tag{5.56}$$

*and*

$$\begin{bmatrix} K_1^2 \Psi_{11} + K_1 \Psi_1 & K_1 K_2 \Psi_{12} \\ \textit{sym.} & K_2^2 \Psi_{22} + K_2 \Psi_2 \end{bmatrix} \in \mathrm{Sym}^+(2). \tag{5.57}$$

*Additionally,*

$$\left\langle \begin{bmatrix} 1 \\ 2 \end{bmatrix}, \begin{bmatrix} K_1^2 \Psi_{11} + K_1 \Psi_1 & K_1 K_2 \Psi_{12} \\ \textit{sym.} & K_2^2 \Psi_{22} + K_2 \Psi_2 \end{bmatrix} \begin{bmatrix} 1 \\ 2 \end{bmatrix} \right\rangle > 0. \tag{5.58}$$

*These sufficient conditions and therefore Hill's inequality are implied by the sufficient conditions for polyconvexity from Theorem 3.1 if the nuances related to positivity vs. non-negativity are set aside. With this caveat in mind, every incompressible polyconvex strain-energy $W$ conforming to Theorem 3.1 automatically satisfies Hill's condition and TSTS-M$^{++}$.*

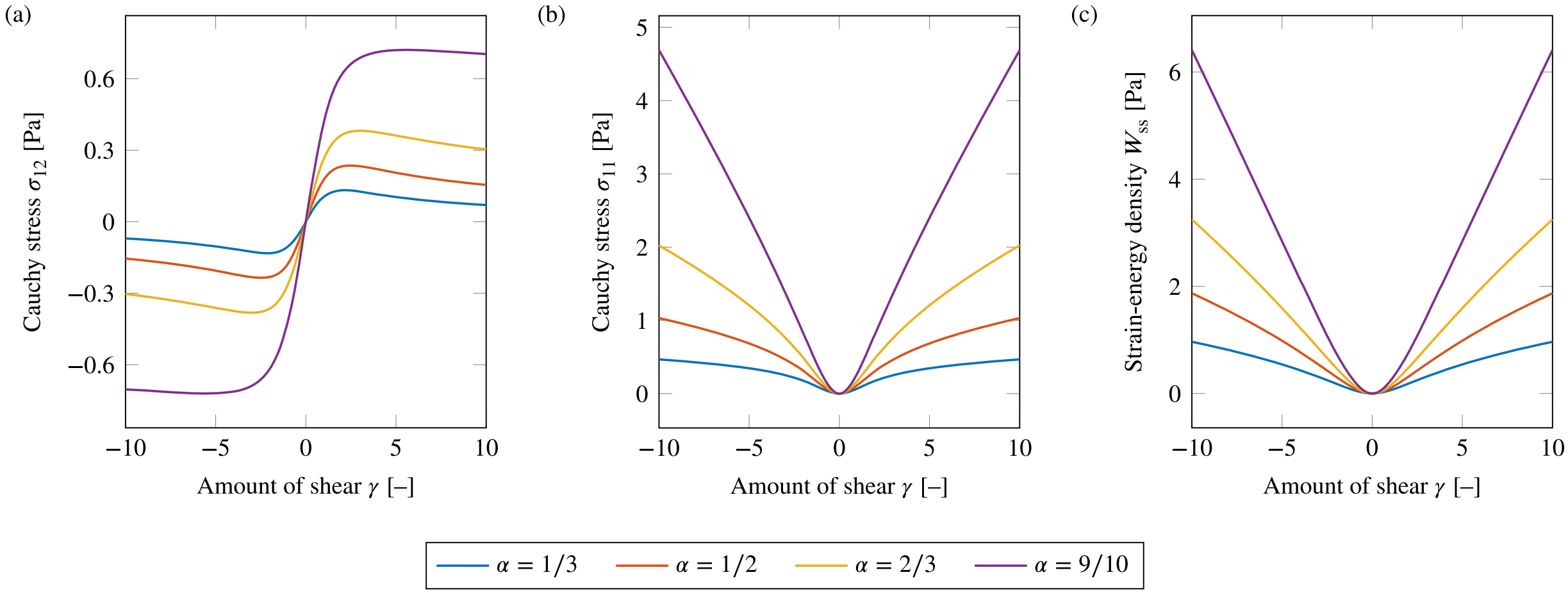

**Fig. 3.** Trajectory of (a) the non-monotonic Cauchy shear stress $\sigma_{12}$, (b) the Cauchy normal stress $\sigma_{11}$, and (c) the strain-energy density $W_{ss}$ during simple shear given the strain-energy function (5.69) for some applied amount of shear $\gamma$. The other normal components not displayed read $\sigma_{22} = \sigma_{33} = -\frac{1}{2}\sigma_{11}$. The legend applies throughout.

### 5.3. *True-shear-stress monotonicity in simple shear*

Here, we give a family of solutions for Challenge (iii), i.e., a strain-energy function $W$ that satisfies TSTS-M$^{++}$, but leads to a non-monotonic true-shear-stress response. This immediately entails that $W$ is not rank-one convex as shown in Proposition (5.13). By simple shear we refer to a motion leading to a constant deformation gradient in the form

$$\mathbf{F} = \mathbb{1} + \gamma \boldsymbol{e}_1 \otimes \boldsymbol{e}_2, \tag{5.59}$$

where $\gamma \in \mathbb{R}$ denotes the amount of shear. In the construction of a valid candidate function, we encounter a non-linear ordinary differential equation which is solved in Lemma 5.15.

**Proposition 5.13.** *Let the strain-energy function $W$ be rank-one convex and continuously differentiable. Then its true-shear-stress response in simple shear is monotonic.*

**Proof.** If $W$ is rank-one convex and continuously differentiable, then

$$\langle \mathbf{S}_1(\overline{\mathbf{F}}) - \mathbf{S}_1(\mathbf{F}), \overline{\mathbf{F}} - \mathbf{F} \rangle \geq 0, \tag{5.60}$$

where $\mathbf{S}_1$ denotes the first Piola-Kirchhoff stress tensors resulting from the deformation gradients $\overline{\mathbf{F}}, \mathbf{F} \in \mathrm{GL}^+(3)$ for which

$$\overline{\mathbf{F}} = \mathbf{F} + \boldsymbol{a} \otimes \boldsymbol{b}, \tag{5.61}$$

cf. (Šilhavý, 1997, Sect. 17.3) and (Ogden, 1997, App. 1).

With $\boldsymbol{\sigma} = \frac{1}{J}\mathbf{S}_1\,\mathbf{F}^{\mathrm{T}}$, cf. (Truesdell and Noll, 1965, Eq. (43 A.3)), and

$$\mathbf{F} = \mathbb{1} + \gamma\, \boldsymbol{e}_1 \otimes \boldsymbol{e}_2 \qquad \text{and} \qquad \overline{\mathbf{F}} = \mathbb{1} + \overline{\gamma}\, \boldsymbol{e}_1 \otimes \boldsymbol{e}_2, \tag{5.62}$$

the inequality (5.60) reduces to

$$\big(\sigma_{12}(\overline{\gamma}) - \sigma_{12}(\gamma)\big)\,(\overline{\gamma} - \gamma) \geq 0, \tag{5.63}$$

i.e., the true-shear-stress response in simple shear is monotonic. □

**Remark 5.14.** *As shown in (Voss et al., 2020), rank-one convexity and in turn polyconvexity ensure uniqueness of solutions in a broader class of spatially heterogeneous deformation modes called 'anti-plane shear' involving only one scalar function in the definition of the motion $\boldsymbol{\varphi}$. This provides further support for rank-one convexity as a sensible constitutive requirement for an idealized elastic response.*

**Lemma 5.15.** *The ordinary non-linear differential equation*

$$x^2\, u\, u'' - \left(x\, u' - \frac{u}{2}\right)^2 = \frac{k\, u^2}{4}, \tag{5.64}$$

*defined for $k \in \mathbb{R}$ over $x > 0$, has the general solution*

$$u(x) = c_2\, x^{c_1} \exp\left(\frac{k+1}{8} \log^2 x\right), \tag{5.65}$$

*where $c_1$ and $c_2$ are arbitrary constants of integration.*

**Proof.** Substituting $v(y) = u(x)$ with $y = \log x$, we have

$$\dot{v} = u' x \qquad \text{and} \qquad \ddot{v} = u'' x^2 + u' x, \tag{5.66}$$

where the prime and dot denote differentiation with respect to $x$ and $y$, respectively. Thus,

$$x^2 u u'' - \left(x u' - \frac{u}{2}\right)^2 = \frac{k u^2}{4} \quad \Longrightarrow \quad v(\ddot{v} - \dot{v}) - \left(\dot{v} - \frac{v}{2}\right)^2 = v\ddot{v} - \dot{v}^2 - \frac{v^2}{4} = \frac{k v^2}{4}. \tag{5.67}$$

Assuming $v \neq 0$, we further have

$$\frac{\ddot{v}}{v} - \left(\frac{\dot{v}}{v}\right)^2 - \frac{k+1}{4} = \frac{\mathrm{d}^2 \log v}{\mathrm{d}y^2} - \frac{k+1}{4} = 0 \quad \Longrightarrow \quad v(y) = c_2 \exp\left(\frac{k+1}{8} y^2 + c_1 y\right). \tag{5.68}$$

Addressing the prior assumption $v \neq 0$, notice that $v$ cannot vanish unless $c_2 = 0$ which solves the differential equation trivially. Substituting back $y = \log x$, we arrive at the desired result for $u$. □

**Proposition 5.16.** *Let*

$$W(\mathbf{F}) = \frac{\|\mathbf{F}\|^\alpha}{(\det \mathbf{F})^{\alpha/3}} \exp\left(\beta \log^2(\det \mathbf{F})\right) + \text{const.}, \tag{5.69}$$

*where $\alpha \in (0,1)$ and $\beta > \frac{1}{8}$. Then the elastic response function derived from the strain-energy function $W$ satisfies TSTS-M$^{++}$ and shows a non-monotonic true-shear-stress trajectory in simple shear.*

**Proof.** From (5.69) with (2.2), we have

$$\Psi(K_1, K_2, K_3) = K_1^\alpha K_3^{-\alpha/3} \exp\left(\beta \log^2 K_3\right) + \text{const.} \tag{5.70}$$

Following (2.5), the elastic response function for $\Psi$ reads

$$\boldsymbol{\sigma} = K_1^\alpha K_3^{-(\alpha/3+1)} \exp(\beta \log^2 K_3)\left(\frac{\alpha}{K_1^2}\mathbf{B} + \left(-\frac{\alpha}{3} + 2\beta \log K_3\right)\mathbb{1}\right), \tag{5.71}$$

satisfying the constraint (2.10) of a stress-free initial configuration.

From Corollary 4.4.1, we have sufficient conditions for TSTS-M$^{++}$. For ease of exposition, we abbreviate

$$u(K_3) = K_3^{-\alpha/3} \exp(\beta \log^2 K_3) > 0. \tag{5.72}$$

With Silvester's criterion, it suffices that

$$\frac{\partial \Psi}{\partial K_1} = \alpha K_1^{\alpha-1} u > 0, \qquad K_1^2 \frac{\partial^2 \Psi}{\partial K_1^2} + K_1 \frac{\partial \Psi}{\partial K_1} = \alpha^2 K_1^\alpha u > 0, \tag{5.73}$$

and

$$(K_1^2 \Psi_{11} + K_1 \Psi_1) K_3^2 \Psi_{33} - \left(K_1 K_3 \Psi_{13} - \frac{1}{2} K_1 \Psi_1\right)^2 = \alpha^2 K_1^{2\alpha}\left(K_3^2 u u'' - \left(K_3 u' - \frac{u}{2}\right)^2\right) > 0, \tag{5.74}$$

where the prime denotes differentiation with respect to $K_3$. The fulfillment of the last condition follows immediately from Lemma 5.15 for $c_1 = -\frac{\alpha}{3}$, $c_2 = 1$, and $k = 8\beta - 1$. This completes the proof of TSTS-M$^{++}$. Although automatically a valid linear-elastic law in the process, we also provide the material constants of the infinitesimal theory with

$$\mu = \alpha\, 3^{\alpha/2-1} > 0, \qquad \kappa = 2\beta\, 3^{\alpha/2} > 0, \qquad \text{and} \qquad \nu = \frac{9\beta - \alpha}{18\beta + \alpha} \in \left(\frac{1}{26}, \frac{1}{2}\right). \tag{5.75}$$

From (5.71), the true-shear-stress response for the simple-shear deformation (5.59) reads

$$\sigma_{12}(\gamma) = \alpha(3 + \gamma^2)^{\alpha/2-1}\gamma. \tag{5.76}$$

Notice that

$$\lim_{\gamma \to \pm\infty} \sigma_{12}(\gamma) = \lim_{\gamma \to \pm\infty} \frac{\gamma^{\alpha-1}}{(1 + 3\gamma^{-2})^{1-\alpha/2}} = 0. \tag{5.77}$$

Since the centrally symmetric, continuously differentiable true-shear-stress response has a positive initial slope due to (5.75), it follows from Rolle's theorem that the trajectory must have a global maximum and minimum, i.e., it is non-monotonic. □

**Remark 5.17.** *From Theorem 5.13, it follows immediately that $W$ cannot be rank-one convex and in turn not polyconvex. Even with $\alpha \geq 1$, $\Psi$ fails to satisfy the sufficient condition $\Psi_{11}\Psi_{33} - \Psi_{13}^2 \geq 0$ from Theorem 3.1 globally. In Fig. 3, we visualize the trajectories of the Cauchy stress components $\sigma_{11}$ and $\sigma_{12}$ as well as the strain-energy density $W_{\mathrm{ss}}$ over the amount of shear $\gamma$ for a variety of $\alpha$. Here, $W_{\mathrm{ss}}(\gamma)$ refers to $W(\mathbf{F})$ evaluated for the simple-shear deformation (5.59). Since the deformation is isochoric, the parameter $\beta$ has no influence on the stress response. As expected, $W_{\mathrm{ss}}(\gamma) = (3 + \gamma^2)^{\alpha/2}$ is not convex in $\gamma$ for $\alpha \in (0,1)$.*

**Remark 5.18.** *One should note that simple shear at large strains is a famously difficult deformation mode to realize experimentally due to the required application of normal surface tractions, cf. (Rivlin, 1997, Sect. 4). In this sense, a material response to simple shear at finite strains is not as physically 'intuitive' as it might appear at first.*

**Remark 5.19.** *A simple example for a merely Cauchy elastic constitutive relation that satisfies TSTS-M$^{++}$, but shows a non-monotonic true-shear-stress response in simple shear, can be found in Hencky's proposal $\boldsymbol{\sigma} = 2\mu \log \mathbf{V} + \lambda\, \mathrm{tr}(\log \mathbf{V})\mathbb{1}$ from 1928.*

## 6. Conclusion

In this contribution, we discuss two constitutive inequalities in the context of isotropic hyperelasticity: polyconvexity and the true-stress-true-strain monotonicity (TSTS-M$^{++}$). We show that it is possible for a polyconvex strain-energy to produce a non-monotonic true-stress response in unconstrained uniaxial extension. Such behavior would be impossible under TSTS-M$^{++}$. Similarly, we constructed a strain-energy function that obeys TSTS-M$^{++}$, but leads to a non-monotonic Cauchy shear stress response in simple shear – a result at odds with polyconvexity. These explicit examples support the notion that neither of the two constitutive inequalities are sufficient by themselves to ensure physically reasonable material behavior for ideal elasticity.

In case of incompressible material behavior, we show that a strain-energy function that satisfies the sufficient conditions for polyconvexity by (Ball, 1976, Theo. 5.2) has a monotonic true-stress response in unconstrained uniaxial extension-compression. Since these conditions are only sufficient, it remains unclear whether or not an incompressible, polyconvex strain-energy function can show a non-monotonic true-stress response in this deformation mode.

In order to construct valid families of strain-energy functions for these questions, we establish sufficient conditions for both polyconvexity and TSTS-M$^{++}$ in terms of a specific set of invariants. Although these conditions share many features, we have so far not been able to find a strain-energy function that satisfies both constitutive inequalities simultaneously. We are however able to construct such candidates in a chain-limited setting. It might be also possible that a valid strain-energy function, that is defined globally, does not exist. To this end, the study of the here derived conditions for polyconvexity and TSTS-M$^{++}$ might be worthwhile as the combination of both seem to be a reasonable constitutive requirement for hyperelasticity.

## CRediT authorship contribution statement

**Maximilian P. Wollner:** Writing - review & editing, Writing - original draft, Visualization, Methodology, Investigation, Formal analysis, Conceptualization; **Gerhard A. Holzapfel:** Writing - review & editing, Supervision, Funding acquisition; **Patrizio Neff:** Writing - review & editing, Writing - original draft, Supervision, Methodology, Investigation, Formal analysis, Conceptualization.

## Data availability

No data was used for the research described in the article.

## Declaration of competing interest

The authors declare that they have no known competing financial interests or personal relationships that could have appeared to influence the work reported in this paper.

## Acknowledgement

Maximilian P. Wollner acknowledges financial support from the European Union's Horizon 2020 program for research and innovation under grant agreement no. 101017523. Maximilian P. Wollner is especially grateful to Karl A. Kalina (Dresden University of Technology) and Dominik K. Klein (Technical University of Darmstadt) for the many critical discussions about polyconvexity at a number of scientific conferences. Furthermore, Maximilian P. Wollner would like to thank Robert J. Martin (University of Duisburg-Essen), Timo Neumeier (University of Augsburg), and David Wiedemann (Technical University of Dortmund) for their explanations regarding polyconvexity in terms of signed singular values.

## Appendix A. Notation

In this work, both the current and reference configuration share the same Cartesian coordinates system with the orthonormal base vectors $(\boldsymbol{e}_i)_{i=1}^3$ and we omit the distinction between covariant and contravariant indices.

First-order and second-order tensors are written in italic and straight bold font, respectively, e.g., $\boldsymbol{a} = a_i \boldsymbol{e}_i$ and $\mathbf{X} = X_{ij} \boldsymbol{e}_i \otimes \boldsymbol{e}_j$. Here, the symbol '⊗' denotes the dyadic product. The second-order identity tensor is written as $\mathbb{1} = \delta_{ij} \boldsymbol{e}_i \otimes \boldsymbol{e}_j$. A single contraction between two tensor is not denoted explicitly, e.g., $\mathbf{XY} = X_{ik} Y_{kj} \boldsymbol{e}_i \otimes \boldsymbol{e}_j$ or $\mathbf{X}\boldsymbol{b} = X_{ik} b_k \boldsymbol{e}_i$. A double contraction between two second-order tensor is defined as $\langle \mathbf{X}, \mathbf{Y} \rangle = \mathrm{tr}(\mathbf{XY}^{\mathrm{T}}) = X_{ij} Y_{ij}$. Similarly, the dot product between two first-order tensor reads $\langle \boldsymbol{a}, \boldsymbol{b} \rangle = a_i b_i$. The operator $\|(\bullet)\|^2 = \langle (\bullet), (\bullet) \rangle$ refers to the Euclidean norm and Frobenius norm for first-order and second-order tensors, respectively. The cofactor of a second-order tensor is denoted by $\mathrm{Cof}\,\mathbf{X} = \det(\mathbf{X})\mathbf{X}^{-\mathrm{T}}$. With $\mathrm{D}_{\mathbf{X}}(\bullet)$ we write the Fréchet derivative of $(\bullet)$ with respect to $\mathbf{X}$, e.g., $\mathrm{D}_{\mathbf{X}}\mathbf{Y} = \frac{\partial Y_{ij}}{\partial X_{kl}} \boldsymbol{e}_i \otimes \boldsymbol{e}_j \otimes \boldsymbol{e}_k \otimes \boldsymbol{e}_l$. Analogously $\mathrm{D}^2_{\mathbf{X}}(\bullet)$ refers to a second-order Fréchet derivative. The double contraction of a fourth-order tensor with a second-order tensor is denoted by a dot, such that $\mathrm{D}_{\mathbf{X}}\mathbf{Y}.\mathbf{Z} = \frac{\partial Y_{ij}}{\partial X_{kl}} Z_{kl} \boldsymbol{e}_i \otimes \boldsymbol{e}_j$.

In this work, all tensors are defined over the real numbers. The set of second-order tensors with positive determinant is defined as the general linear group $\mathrm{GL}^+(n) = \{\mathbf{X} \in \mathbb{R}^{n\times n} \mid \det \mathbf{X} > 0\}$, while the set of symmetric second-order tensor is denoted as $\mathrm{Sym}(n) = \{\mathbf{X} \in \mathbb{R}^{n\times n} \mid \mathbf{X} = \mathbf{X}^{\mathrm{T}}\}$. We also introduce the set of symmetric, positive semi-definite and definite second-order tensors with $\mathrm{Sym}^+(n) = \{\mathbf{X} \in \mathrm{Sym}(n) \mid \langle \mathbf{X}\boldsymbol{a}, \boldsymbol{a} \rangle \geq 0\ \forall \boldsymbol{a} \in \mathbb{R}^n\}$ and $\mathrm{Sym}^{++}(n) = \{\mathbf{X} \in \mathrm{Sym}(n) \mid \langle \mathbf{X}\boldsymbol{a}, \boldsymbol{a} \rangle > 0\ \forall \boldsymbol{a} \in \mathbb{R}^n \setminus \{\mathbf{0}\}\}$, respectively. The set of all positive real numbers is denoted by $\mathbb{R}^+$.

All quantities related to stress and energy density per unit volume are measured in unit Pa without explicit mention. The notational differentiation between a function and its output is omitted at times to avoid the introduction of new symbols. Special exceptions are $\widehat{W}(\log \mathbf{V})$ and $\widehat{\sigma}(\log \mathbf{V})$, where the parametrization in terms of the Hencky strain $\log \mathbf{V}$ is made explicitly with an overset hat.